\documentclass[12pt]{article}
\pdfoutput=1
\usepackage{graphicx}
\usepackage[round,longnamesfirst]{natbib}
\bibpunct[]{(}{)}{;}{a}{}{,}
\usepackage[noend]{algorithmic}
\usepackage{algorithm}

\usepackage{csquotes,amsmath, amssymb, amsbsy,array,enumitem,cancel,url,rotating,doi,subfig}
\usepackage{color}

\DeclareMathOperator{\E}{E}

\DeclareMathOperator{\logit}{logit}

\DeclareMathOperator{\RandomChoose}{RandomChoose}
\DeclareMathOperator{\0}{\boldsymbol{0}}

\DeclareMathOperator{\Uniform}{Uniform}
\DeclareMathOperator{\Prob}{Pr}

\newcommand{\newtxt}[1]{#1}
\newcommand{\netsymbol}{y}
\newcommand{\Netsymbol}{Y}
\newcommand{\y}{\netsymbol}
\newcommand{\Y}{\Netsymbol}
\newcommand{\yv}{\boldsymbol{\netsymbol}}
\newcommand{\Yv}{\boldsymbol{\Netsymbol}}
\newcommand{\covariatesymbol}{x}
\newcommand{\x}{\covariatesymbol}
\newcommand{\xv}{\boldsymbol{\covariatesymbol}}
\newcommand{\weightsymbol}{w}
\newcommand{\w}{\weightsymbol}
\newcommand{\xs}{\mathbb{X}}
\newcommand{\actors}{{N}}
\newcommand{\egos}{\actors_{\text{e}}}
\newcommand{\alters}{\actors_{\text{a}}}
\newcommand{\actorsnot}[1]{\actors\setsub\left\{#1\right\}}
\newcommand{\distuples}[1]{\actors^{#1\ne}}
\newcommand{\egodistuples}[1]{\egos^{#1\ne}}
\newcommand{\altdistuples}[1]{\alters^{#1\ne}}
\newcommand{\nactors}{ n }
\newcommand{\npar}{ p }
\newcommand{\paramsymbol}{\theta}
\newcommand{\param}{\paramsymbol}
\newcommand{\paramv}{\boldsymbol{\paramsymbol}}
\newcommand{\curvpars}{\boldsymbol{\Theta}}
\newcommand{\natcurvpars}{\curvpars_{\text{N}}}
\newcommand{\genstatsymbol}{g}
\newcommand{\genstats}{\boldsymbol{\genstatsymbol}}
\newcommand{\genstat}[1]{\genstatsymbol_{\text{#1}}}
\newcommand{\dyadvals}{\mathbb{S}}
\newcommand{\pref}[1][]{\myrel{\ensuremath{\,\,#1}}{\succ}}
\newcommand{\npref}[1][]{\myrel[-0.2]{\ensuremath{#1}}{\nsucc}}
\newcommand{\indiff}[1][]{\myrel{\ensuremath{#1}}{\cong}}
\newcommand{\promote}[2]{\Delta_{#1,#2}^\nearrow}
\newcommand{\ipromotej}{\promote{i}{j}}
\newcommand{\promotev}[2]{\boldsymbol{\Delta}_{#1,#2}^\nearrow}
\newcommand{\ipromotejv}{\promotev{i}{j}}
\newcommand{\setsub}{\backslash}
\newcommand{\dysY}{\mathbb{Y}}
\newcommand{\netsY}{\mathcal{Y}}
\newcommand{\Pteg}{\Prob_{\paramv;\genstats}}
\newcommand{\Eteg}{\E_{\paramv;\genstats}}
\newcommand{\Ptegx}{\Prob_{\paramv;\genstats,\xv}}
\newcommand{\normc}{\kappa}
\newcommand{\ceg}{\normc_{\genstats}}
\newcommand{\cegx}{\normc_{\genstats,\xv}}
\newcommand{\ilogit}{\logit^{-1}}
\newcommand{\reals}{\mathbb{R}}
\newcommand{\naturals}{\mathbb{N}}
\renewcommand{\ij}{{i,j}}
\newcommand{\ynetsY}{{\yv\in\netsY}}
\newcommand{\ypnetsY}{{\yv'\in\netsY}}
\newcommand{\sij}{_{\ij}}
\newcommand{\sijk}{_{i,j,k}}
\newcommand{\sik}{_{i,k}}
\renewcommand{\l}{l}
\newcommand{\sil}{_{i,\l}}
\newcommand{\yij}{\y\sij}
\newcommand{\Yy}{\Yv=\yv}
\newcommand{\jplus}{j^{+}}
\newcommand{\EN}[3]{\left#1 #3 \right#2}
\newcommand{\en}[3]{#1 #3 #2}
\newcommand{\yvat}[1]{\yv^{t#1}}
\newcommand{\Yvat}[1]{\Yv^{t#1}}
\newcommand{\myexp}[1]{\exp\mathchoice{\left(#1\right)}{(#1)}{(#1)}{(#1)}}
\newcommand{\egopref}[4]{#1_{#2:\,#3\pref #4}}
\newcommand{\ypref}[3]{\egopref{\y}{#1}{#2}{#3}}
\newcommand{\yvpref}[3]{\egopref{\yv}{#1}{#2}{#3}}
\newcommand{\egoswapr}[4]{#1^{#2:\,#3\rightleftarrows#4}}
\newcommand{\pkg}[1]{\texttt{#1}}
\newcommand{\proglang}[1]{\textsf{#1}}
\newcommand{\fromthru}[2]{\left\{#1\,..\,#2\right\}}
\newcommand{\innerprod}[2]{{#1}\cdot{#2}}
\newcommand{\centercol}[1]{\multicolumn{1}{c}{#1}}
\newcommand{\coef}[3]{$#1~(#2)^{#3}$}
\newcommand{\xemself}{himself or herself}
\newcommand{\aA}{\text{A}}
\newcommand{\aB}{\text{B}}
\newcommand{\aC}{\text{C}}
\newcommand{\aD}{\text{D}}
\newcommand{\blankcell}{\square}
\newcommand{\xe}{he or she}
\newcommand{\xer}{his or her}
\newcommand{\xem}{him or her}
\newcommand{\yex}{\yv}
\newcommand{\yexbc}{\egoswapr{\yv}{\aA}{\aB}{\aC}}
\newcommand{\yexbd}{\egoswapr{\yv}{\aA}{\aB}{\aD}}
\newcommand{\yexw}{\yv^{\star}}
\newcommand{\abs}[1]{\left\lvert#1\right\rvert}
\makeatletter
\newcommand{\myrel}[3][.3]{\binrel@{#3}%
  \binrel@@{\mathop{\kern\z@#3}\limits^{\vbox to #1\ex@{\kern-\tw@\ex@
\hbox{\scriptsize #2}\vss}}}}
\makeatother
\newcommand{\Yyvat}[1]{\Yvat{#1}=\yvat{#1}}

\title{Exponential-Family Random Graph Models for Rank-Order Relational Data}
\date{}
\author{Pavel N. Krivitsky\footnote{
    University of Wollongong, Wollongong, New South Wales, Australia.  
    Correspondence should be sent to: \href{mailto:pavel@uow.edu.au}{\nolinkurl{pavel@uow.edu.au}}.  
  }
\and 
Carter T. Butts\footnote{
    University of California, Irvine, California, USA.  
  }
}

\begin{document}
\maketitle
\begin{abstract}
Rank-order relational data, in which each actor ranks the others according to some criterion, often arise from sociometric measurements of judgment (e.g., self-reported interpersonal interaction) or preference (e.g., relative liking). We propose a class of exponential-family models for rank-order relational data and derive a new class of sufficient statistics for such data, which assume no more than within-subject ordinal properties. Application of MCMC MLE to this family allows us to estimate effects for a variety of plausible mechanisms governing rank structure in cross-sectional context, and to model the evolution of such structures over time. We apply this framework to model the evolution of relative liking judgments in an acquaintance process, and to model recall of relative volume of interpersonal interaction among members of a technology education program.
\end{abstract}

\shortcites{harris2003nls,morse.et.al:jp:1974,berger.et.al:bk:1977,anderson.et.al:jpsp:2006,handcock2010epf,snijders2006nse,hunter2008epf,bernard1984pia}

\section{Introduction}
Rank-order sociometric data---in which each actor in a network ranks the others according to some criterion---have a long history in the social sciences.
\newtxt{Among many, many other instances, \citet{sampson1968npc} famously asked each of 18 novitiates in a monastery to rank his three most liked novitiates among the other 17; \citet{newcomb1961ap} measured evolving rankings of each other by 17 men living in fraternity-style housing over the course of a semester; and, more recently, Wave I of National Longitudinal Study of Adolescent Health asked high school students to list, in order, up to 5 male and up to 5 female friends \citep{harris2003nls}.}
While many
network processes (e.g., diffusion, brokerage, exchange) are only
sensibly posited for networks with categorical or ratio scale
relationship states, many others---particularly those involving
personal preferences (e.g., liking, advice-seeking)---are much more
readily represented ordinally, and, indeed, may not even have
interval, ratio, or categorical meaning across raters.  This last can be true even when data is not collected in an explicitly ordinal fashion.  For instance, \citet{johnson.et.al:jms:2003} asked personnel in an isolated environment (the Amundsen--Scott South Pole Station) to rate their degree of interaction with each other on a 0--10 scale, with 0 indicating no interaction and 10 indicating a ``great deal'' of interaction.  While it is reasonable to assume that such ratings are ordinally coherent within rater (e.g., if Bob rates Sally below Jill, then Bob regards himself as interacting more with Jill than with Sally), such ratings cannot be compared \emph{across} raters: if Bob rates his interaction with Jill at 4, and Sally rates her interaction with Jill at 6, we have no basis for concluding that Sally's interaction with Jill is stronger than Bob's.  Such interaction rating data is thus ``local'' to the rater, and must be analyzed in a manner that avoids cross-rater comparisons.

As the above suggests, much valued sociometric data based on individual ratings or judgments poses issues akin to explicit rank-order data.  Indeed, is common practice in statistics to reduce data with unknown or problematic distributional form to ranks for nonparametric analyses such as Kruskal--Wallis and Mann--Whitney procedures, and similar use may be made of ranks in networks with valued ties as a compromise between analyzing tie values as they are and dichotomizing them \citep[:32--34, for example]{newcomb1961ap}.  Where there is no objective basis for equating or ordering tie values across raters, such treatment of valued network data reduces to the rank-order data case.

The most common approach taken to analyzing rank-valued network data in current practice is to dichotomize ranks into binary ties, defining a tie to be present if a given ego had ranked a given alter above a certain cut-off and absent otherwise. Many methods of dichotomizing have been proposed.  For instance, cut-offs have been set at a particular rank (e.g., top 5) \citep[, for example]{breiger1975acr, white1976ssm,  arabie1978cbh, wasserman1980asn, pattison1982asm, harris2003nls}; at a particular quantile (e.g., top 50\%) \citep{krackhardt2007hvs}; or have been found adaptively \citep{doreian1996bhb}. Another common approach is to focus on rank correlations and on treating ranks on an additive scale \citep{newcomb1956pia,nakao1993las}.

These approaches come with significant limitations. Dichotomizing ties
requires a threshold point to be selected, inevitably discarding information and possibly introducing biases \citep{thomas2011vtt},
while methods like rank correlation are limited to simple comparisons
and cannot, for example, be used to examine the strength of one social
factor after controlling for the effects of another.  \emph{More importantly, these techniques implicitly assume that tie values can be equated across raters, an assumption that is often unjustified.}  When the presence of an $(i,j)$ tie has a different empirical meaning than the presence of a $(k,j)$ tie, conventional network analytic techniques (e.g., centrality indices) may prove misleading.

Modeling frameworks explicitly designed for rank-order data would
address these limitations, but to date 
work on model-based approaches to rank-order network data has been very limited.
\citet{gormley2008evb}, for instance, use a
generalization of the Plackett--Luce model \citep{plackett1975ap} in a
latent position framework to model what can be viewed as a bipartite
rank-order network of affiliations from voters to candidates in Irish
proportional representation through the single transferable vote
elections.  Null models for comparison of rank-order (or otherwise valued) data structures were developed by \citet{hubert:bk:1987}, and model-based extensions of this approach for comparison of multiple structures have been introduced by \citet{butts2007pmr}. (See also related work on null hypothesis testing in a network regression context, e.g., that of \citet{krackhardt:sn:1987} and \citet{dekker.et.al:p:2007}.)  This latter work is focused on modeling degrees of correspondence \emph{between} relational structures, and does not attempt to model the internal properties of rank-valued networks themselves.  This second problem is the focus of the present paper.

For modeling of internal network structure, exponential-family random graph (ERG) or $p^*$ models 
\citep{holland1981efp,wasserman1996lml,robins1999lml} are the currently favored approach.  Models parameterized in this way have been applied to social network data in a variety of contexts, including dichotomized rank-order data \citep{krackhardt2007hvs,goodreau2008bff}; used in this latter capacity, they inherit the difficulties with dichotomizing noted above. \citet{robins1999lml} were the first to introduce a systematic treatment of ERGMs for categorically valued network data, along with procedures for approximate inference using pseudo-likelihood estimation.  The model families they propose can be used when edge values are ordinal in an absolute sense, but assume (1) that values can be meaningfully equated across raters, and (2) that there is a well-defined zero-value indicating the absence of a tie, that is qualitatively different from other possible tie values (and, per (1), equivalent across raters).  Building on this work, \citet{krivitsky2012erg} formulated a generalized framework for exponential-family models on networks whose ties have values (categorical or otherwise) and introduced Markov chain Monte Carlo (MCMC) methods for simulation and maximum-likelihood inference in this more general case.  This formulation provides a basis for generalizing to models of the kind we consider here, but retains the assumption of ``absolute'' edge values that have a constant meaning across raters.

In this paper, we develop ERG models for ``locally'' ordinal relational data in which ratings cannot be directly compared across subjects; we focus on the foundational case of complete rankings, but introduce model terms that can be used with more general models (e.g., for partial orders). In Section~\ref{sec:framework}, we discuss representation of ordinal relational data and introduce the probabilistic framework for exponential-family models for them, and in Section~\ref{sec:terms}, we describe statistics that can be used to model common network properties within this framework. We discuss issues involved in implementation and statistical inference on these models in Section~\ref{sec:inference}. Two applications of this framework are demonstrated in Section~\ref{sec:examples}.

\section{\label{sec:framework}Exponential-Family Framework for Ordinal Relational Data}
\subsection{Actors, Rankings, and Comparisons}
We begin this section by defining notation for representation of
ordinal relational data and by establishing basic principles for using
such data in a manner that respects its intrinsic measurement
properties.

\newtxt{Consider a set of $\nactors$ actors of interest, $\actors$, whom we index as
  $\actors=\{\aA,\aB,\aC,\dotsc\}$.  Each actor in $\actors$ will be a rater in our
  network of interest; except as noted otherwise, the objects being rated by each rater are the other members of $\actors$ (though we will consider other possibilities in Section~\ref{sec:bipartite}).  For various purposes, it will be helpful to have specific notation to refer to the set of possible $p$-tuples of distinct actors in $\actors$; we refer to this as the $p$th \enquote{distinct Cartesian power} of $\actors$.
  Recall that the $p$th (ordinary) Cartesian
  power of a set $\actors$ is the set of all ordered $p$-tuples of
  individuals from $\actors$. For example, if
  $\actors=\{\aA,\aB,\aC\}$, then second Cartesian power of $\actors$ is 
  $\actors^2=\{(\aA,\aA), (\aA,\aB),
  (\aA,\aC), (\aB,\aA), (\aB,\aB), (\aB,\aC), (\aC,\aA), (\aC,\aB),
  (\aC,\aC)\}$. By analogy, we denote the $p$th distinct Cartesian power of $\actors$ (the set containing all $p$-tuples of $\actors$ whose elements do not repeat) by $\actors^{p\ne}\subset\actors^p$.\footnote{In formal notation, this can be written as
    $\distuples{p}\equiv\{\boldsymbol{v}\in \actors^p: \forall_{i\in
      \fromthru{1}{p}}\forall_{j\in \fromthru{1}{p}} i\ne j \implies
    v_i\ne v_j\}.$}  Following our example, then, the second \emph{distinct} Cartesian power of $\actors$ is given by
  \break$\actors^{2\ne}=\{\cancel{(\aA,\aA)}, (\aA,\aB), (\aA,\aC),
  (\aB,\aA), \cancel{(\aB,\aB)}, (\aB,\aC), (\aC,\aA), (\aC,\aB),
  \cancel{(\aC,\aC)}\}$. }

\newtxt{As discussed in the introduction, our data consist of
  observations in which each actor (\emph{ego}) in the network $i\in\actors$
  provides some ranking or ordering of the other actors (\emph{alters}),
  and, possibly, of \xemself.  In other words, each actor $i$
  defines an ordinal relation $\pref[i]$ over set
  $\actors$. This relation could represent \enquote{preferred to},
  \enquote{interacted more with than}, \enquote{judged to be taller
    than}, or any other judgment of interest.  Importantly, we note that for two
  egos $i$ and $j$, $\pref[i]$ need not equal $\pref[j]$, and the ratings of some
  alter $k$ by $i$ and $j$ cannot be compared directly; we may, however, meaningfully
  ask whether, e.g. $(k \pref[i] \l) = (k \pref[j] \l)$---whether $i$ and $j$'s
  rankings of $k$ and $\l$ are concordant, and our modeling framework is founded
  on exactly these distinctions.}

\begin{figure}
  \centerline{\includegraphics[width=.8\textwidth]{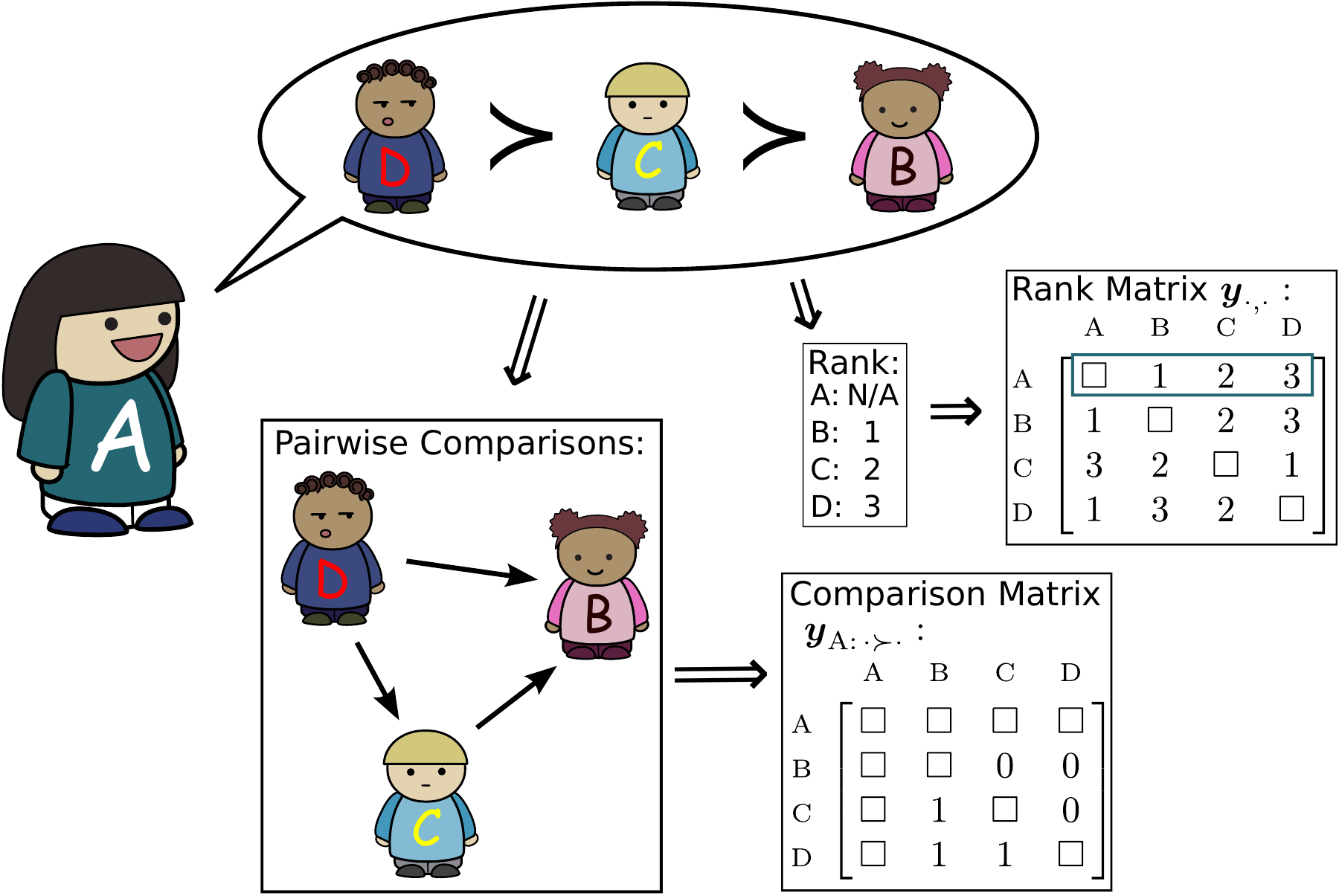}}
  \caption{\label{fig:representations} \newtxt{ Ego~$\aA$'s report
    regarding her ranking of $\aB$, $\aC$, and $\aD$ can be encoded as a rank ordering or
    as pairwise comparisons.  Here, Ego~$\aA$'s response ranks $\aD$ highest, then
    $\aC$, then $\aB$. We may encode this report by assigning a rank of 1 to
    $\aB$, 2 to $\aC$, and 3 to $\aD$ resulting in a row of rank matrix $\yv_{\cdot,\cdot}$; or we may consider all pairwise
    comparisons implied by $\aA$: that $\aD$ is ranked over $\aC$, $\aD$ is
    ranked over $\aB$, and $\aC$ is ranked over $\aB$, and the opposite does
    not hold, resulting in a binary matrix of comparison indicators
    $\yvpref{\aA}{\cdot}{\cdot}$. Here, boxes
    ($\blankcell$) denote matrix entries that are unobservable and/or
    meaningless: in this case, these include $\aA$'s row and column entries in the comparison matrix and the diagonal of the rank matrix, because $\aA$ is not permitted to rank herself among
    the others; and the diagonal of the comparison matrix, because it is meaningless to compare an alter to \xemself.}}
\end{figure}

\newtxt{A simple example of a ranking structure is provided in Figure~\ref{fig:representations}, in which actor $\aA$ (ego) ranks (alter) $\aD$ above $\aB$ and $\aC$ and $\aC$ above $\aB$.  In our above notation, this corresponds to $\aD\pref[\aA]\aC$, $\aC\pref[\aA]\aB$, and $\aD\pref[\aA]\aB$.  The presence of corresponding structures associated with Egos $\aB$, $\aC$, and $\aD$ (not illustrated, but shown in the rank matrix of Figure~\ref{fig:representations}, right) results in a complete ordinal network.}

\newtxt{In general, we make few assumptions regarding
  $\pref[i]$. Assuming the ego may not include self in the ranking, we
  require that for all ego--alter--alter triples $(i,j,k) \in
  \distuples{3}$, $i$ reports either $j\pref[i] k$ or $j\npref[i]
  k$. With the additional assumption of transitivity ($j\pref[i]k\land
  k\pref[i]\l \implies j\pref[i]\l$), the above formulation leads to
  representing a partial ordering of alters by each ego. If
  incomparability is also transitive ($j\npref[i]k\land k\npref[i]\l
  \implies j\npref[i]\l$), a weak ordering results, which may be used
  to represent rank data in which \enquote{ties} are allowed.
  Finally, a further constraint on $\pref[i]$ that $j\npref[i]k \implies
  k\pref[i]j$ results in a complete ordering, which may arise when an
  ego is forced to rank all of the alters with no equal ranks
  permitted. This is an important special case, and we consider it
  here in more detail.  }

\newtxt{We will, furthermore, focus on the case where the ego does not
  report a ranking for self and where the set of egos is the same as
  the set of alters (e.g., people ranking other people who rank them
  in turn, as opposed to e.g. consumers ranking brands or other objects), so relation
  $j\pref[i]k$ is meaningful only for $(i,j,k) \in \distuples{3}$. If
  an ego is permitted to rank itself among the others, it is also
  possible for $j$ or $k$ (but not both) to equal $i$: $(i,j,k) \in
  \actors\times\distuples{2}$, and if we call the set of people in the
  people-ranking-objects scenario $\egos$ and the set of objects
  $\alters$, then $(i,j,k) \in \egos\times\altdistuples{2}$.  This
  last case is discussed further in Section~\ref{sec:bipartite}.
}

\newtxt{Rankings are not assumed to be comparable across egos:
  for each ego $i$, we may only say that one alter is \enquote{$\pref[i]$}
  another (or that this does not hold). To concisely represent
    this fundamental operation in a specific ordinal network $\yv$, we define an indicator
\begin{equation}
\ypref{i}{j}{k}\equiv\begin{cases}
  1 & \text{if $j\pref[i]k$ i.e.,  $i$ ranks $j$ above $k$;} \\
  0 & \text{otherwise.}
\end{cases}\label{eq:egopref}
\end{equation}
This represents the basic distinction that can be unequivocally made
from locally ordinal data. As we shall see in Section~\ref{sec:terms}, being limited to such
statements does not prevent us from specifying a very rich class of
models.}

\subsection{Representations of Ordinal Networks}

\newtxt{We make use of two numerical representations of the observed
  networks of rankings, which we illustrate on toy networks with
  $\nactors=4$: a network of complete orderings $\yex$ in
  Figure~\ref{fig:toy-orig} and a network of idiosyncratic ranking
  structures $\yexw$ in
  Figure~\ref{fig:toy-weird}. Figure~\ref{fig:toy} also shows
  representations of two perturbations (\subref{fig:toy-ABC} and \subref{fig:toy-ABD}) of $\yex$. We make use of them
  in subsequent sections on model terms and interpretations.}
\begin{figure}
\centering
\newtxt{
\subfloat[$\yex$: original network]{\label{fig:toy-orig}
\fbox{{\small
\begin{tabular}{cccc}
\multicolumn{2}{c}{$\yex$ Ranking:} & \multicolumn{2}{c}{$\yex_{\cdot,\cdot}$} \\
\multicolumn{2}{c}{$
\begin{matrix}
    \aA: & \aD\pref\aC\pref\aB \\
    \aB: & \aD\pref\aC\pref\aA \\
    \aC: & \aA\pref\aB\pref\aD \\
    \aD: & \aB\pref\aC\pref\aA \\
\end{matrix}
$}
&
\multicolumn{2}{c}{$ 
\begin{bmatrix}
\blankcell & 1 & 2 & 3 \\
1 & \blankcell & 2 & 3 \\
3 & 2 & \blankcell & 1 \\
1 & 3 & 2 & \blankcell \\
\end{bmatrix} $}
\\
$\yvpref{ \aA }{\cdot}{\cdot}$ & $\yvpref{ \aB }{\cdot}{\cdot}$ & $\yvpref{ \aC }{\cdot}{\cdot}$ & $\yvpref{ \aD }{\cdot}{\cdot}$ \\
  $ 
\begin{bmatrix}
\blankcell & \blankcell & \blankcell & \blankcell \\
\blankcell & \blankcell & 0 & 0 \\
\blankcell & 1 & \blankcell & 0 \\
\blankcell & 1 & 1 & \blankcell \\
\end{bmatrix} $ & $ 
\begin{bmatrix}
\blankcell & \blankcell & 0 & 0 \\
\blankcell & \blankcell & \blankcell & \blankcell \\
1 & \blankcell & \blankcell & 0 \\
1 & \blankcell & 1 & \blankcell \\
\end{bmatrix} $ & $ 
\begin{bmatrix}
\blankcell & 1 & \blankcell & 1 \\
0 & \blankcell & \blankcell & 1 \\
\blankcell & \blankcell & \blankcell & \blankcell \\
0 & 0 & \blankcell & \blankcell \\
\end{bmatrix} $ & $ 
\begin{bmatrix}
\blankcell & 0 & 0 & \blankcell \\
1 & \blankcell & 1 & \blankcell \\
1 & 0 & \blankcell & \blankcell \\
\blankcell & \blankcell & \blankcell & \blankcell \\
\end{bmatrix} $ 
\end{tabular}
}}
}
}\\
\newtxt{
\subfloat[$\yexbc$: Ego~$\aA$ swapped $\aB$ and $\aC$]{\label{fig:toy-ABC}
\fbox{{\small
\begin{tabular}{cc}
\multicolumn{2}{c}{$\yexbc$ Ranking:} \\
\multicolumn{2}{c}{$
\begin{matrix}
    \aA: & \aD\pref\aB\pref\aC \\
    \aB: & \aD\pref\aC\pref\aA \\
    \aC: & \aA\pref\aB\pref\aD \\
    \aD: & \aB\pref\aC\pref\aA \\
\end{matrix}
$}
\\
 $\yexbc_{\cdot,\cdot}$ &  $\egopref{\yexbc}{ \aA }{\cdot}{\cdot}$ \\
$ 
\begin{bmatrix}
\blankcell & 2 & 1 & 3 \\
1 & \blankcell & 2 & 3 \\
3 & 2 & \blankcell & 1 \\
1 & 3 & 2 & \blankcell \\
\end{bmatrix} $
& $ 
\begin{bmatrix}
\blankcell & \blankcell & \blankcell & \blankcell \\
\blankcell & \blankcell & 1 & 0 \\
\blankcell & 0 & \blankcell & 0 \\
\blankcell & 1 & 1 & \blankcell \\
\end{bmatrix} $ 
\end{tabular}
}}
}
}
\qquad
\newtxt{
\subfloat[$\yexbd$: Ego~$\aA$ swapped $\aB$ and $\aD$]{\label{fig:toy-ABD}
\fbox{{\small
\begin{tabular}{cc}
\multicolumn{2}{c}{$\yexbd$ Ranking:} \\ 
\multicolumn{2}{c}{$
\begin{matrix}
    \aA: & \aB\pref\aC\pref\aD \\
    \aB: & \aD\pref\aC\pref\aA \\
    \aC: & \aA\pref\aB\pref\aD \\
    \aD: & \aB\pref\aC\pref\aA \\
\end{matrix}
$}
\\
 $\yexbd_{\cdot,\cdot}$ & $\egopref{\yexbd}{ \aA }{\cdot}{\cdot}$ \\ 
$ 
\begin{bmatrix}
\blankcell & 3 & 2 & 1 \\
1 & \blankcell & 2 & 3 \\
3 & 2 & \blankcell & 1 \\
1 & 3 & 2 & \blankcell \\
\end{bmatrix} $
& $ 
\begin{bmatrix}
\blankcell & \blankcell & \blankcell & \blankcell \\
\blankcell & \blankcell & 1 & 1 \\
\blankcell & 0 & \blankcell & 1 \\
\blankcell & 0 & 0 & \blankcell \\
\end{bmatrix} $
\end{tabular}
}}
}
}
\caption{
 \label{fig:toy}
\newtxt{Representations of complete rankings
    $\yv$ and illustration of $\egoswapr{\yv}{i}{j}{k}$ (defined in
    Section~\ref{sec:promotion}) for $\nactors=4$. Here, boxes
    ($\blankcell$) denote matrix entries that are unobservable and/or
    meaningless. In $\yexbc$
    \protect\subref{fig:toy-ABC} and $\yexbd$
    \protect\subref{fig:toy-ABD}, pairwise comparison matrices
    $\yvpref{i}{\cdot}{\cdot}$ for $i$ other than $\aA$ are unchanged
     from those of $\yv$ \protect\subref{fig:toy-orig}.
}
}
\end{figure}
\begin{figure}
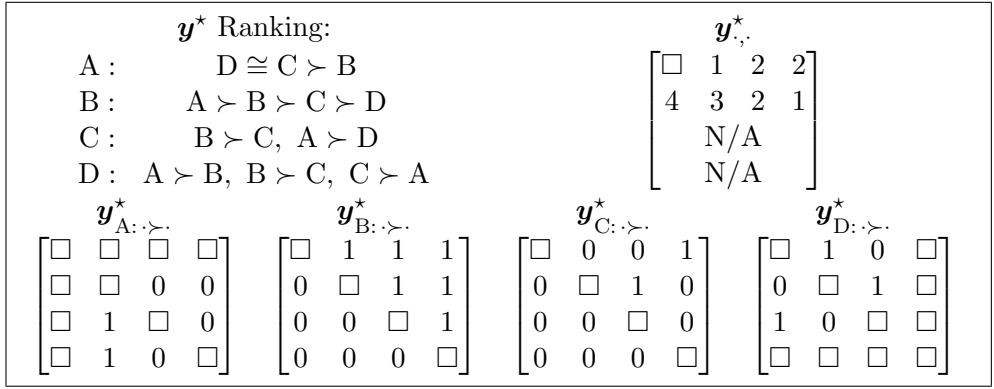

\newtxt{
\begin{center}
\fbox{{\small
\begin{tabular}{cccc}
\multicolumn{2}{c}{$\yexw$ Ranking:} & \multicolumn{2}{c}{$\yexw_{\cdot,\cdot}$}\\
\multicolumn{2}{c}{
$
\begin{matrix}
    \aA: & \aD\indiff\aC\pref\aB \\
    \aB: & \aA\pref\aB\pref\aC\pref\aD \\
    \aC: & \aB\pref\aC,~\aA\pref\aD \\
    \aD: & \aA\pref\aB,~\aB\pref\aC,~\aC\pref\aA \\
\end{matrix}
$
} & \multicolumn{2}{c}{
$ 
\begin{bmatrix}
\blankcell & 1 & 2 & 2 \\
4 & 3 & 2 & 1 \\
\multicolumn{4}{c}{\text{N/A}} \\
\multicolumn{4}{c}{\text{N/A}} \\
\end{bmatrix}
$
}
\\
$\egopref{\yexw}{ \aA }{\cdot}{\cdot}$ & $\egopref{\yexw}{ \aB }{\cdot}{\cdot}$ & $\egopref{\yexw}{ \aC }{\cdot}{\cdot}$ & $\egopref{\yexw}{ \aD }{\cdot}{\cdot}$ 
\\
$ 
\begin{bmatrix}
\blankcell & \blankcell & \blankcell & \blankcell \\
\blankcell & \blankcell & 0 & 0 \\
\blankcell & 1 & \blankcell & 0 \\
\blankcell & 1 & 0 & \blankcell \\
\end{bmatrix} $ & $ 
\begin{bmatrix}
\blankcell & 1 & 1 & 1 \\
0 & \blankcell & 1 & 1 \\
0 & 0 & \blankcell & 1 \\
0 & 0 & 0 & \blankcell \\
\end{bmatrix} $ & $ 
\begin{bmatrix}
\blankcell & 0 & 0 & 1 \\
0 & \blankcell & 1 & 0 \\
0 & 0 & \blankcell & 0 \\
0 & 0 & 0 & \blankcell \\
\end{bmatrix} $ & $ 
\begin{bmatrix}
\blankcell & 1 & 0 & \blankcell \\
0 & \blankcell & 1 & \blankcell \\
1 & 0 & \blankcell & \blankcell \\ 
\blankcell & \blankcell & \blankcell & \blankcell \\
\end{bmatrix} $\\
\end{tabular}
}}
\end{center}
}

\caption{\label{fig:toy-weird}\newtxt{Representations of
    idiosyncratic ordering structures for $\nactors=4$. Here, boxes
    ($\blankcell$) denote matrix entries that are unobservable and/or
    meaningless. Ego~$\aA$ does not rank $\aC$ or $\aD$ over one another but ranks
    both above $\aB$, for a partial ordering; Ego~$\aB$ is permitted to rank
    self along with the others; Ego~$\aC$ is permitted to rank self, but
    does not establish a weak order, since sets $\{\aA,\aD\}$ and
    $\{\aB,\aC\}$ are incomparable; and Ego~$\aD$ reports comparisons that violate
    transitivity. Notice that not all these can be represented using
    a rank matrix $\yexw_{\cdot,\cdot}$, but all can be represented using
    the comparison operation $\ypref{\cdot}{\cdot}{\cdot}$ and sample
    space constraints.}}
\end{figure}

\newtxt{Firstly, a network of complete or a weak ordering $\yv$ may be
  encoded simply as a row in an $\nactors\times\nactors$-matrix of
  ranks that we denote $\yv_{\cdot,\cdot}$, with $\yij$ being the rank by ego $i$ of alter $j$. Consider the ranking
  structure in network $\yv$ in Figure~\ref{fig:toy-orig}. As
  illustrated in Figure~\ref{fig:representations}, the ranking
  reported by Ego $\aA$ can be expressed by assigning the highest
  possible rank $\nactors-1=3$ to $\aD$, the highest-ranked alter,
  $\nactors-2=2$, to $\aC$, the alter with the next highest ranking,
  and $1$ to $\aB$, the actor with the lowest ranking. This
  representation is slightly misleading, in that these ranks are not
  comparable across egos (rows). If the ordering is weak (i.e., there
  are ties), alters may share ranks, as with Ego $\aA$ ranking Alters $\aB$ and $\aC$ equally in the
  idiosyncratic network $\yexw$ in Figure~\ref{fig:toy-weird}. The
  diagonal of $\yv_{\cdot,\cdot}$ is undefined, unless the egos also
  rank themselves in the data as Egos~$\aB$ and~$\aC$ do in Figure~\ref{fig:toy-weird}. }

\newtxt{Secondly, we may represent the comparisons reported by an ego $i$,
  or implied by $i$'s ranking, in a binary
  $\nactors\times\nactors$-matrix of pairwise comparisons, which we
  denote $\yvpref{i}{\cdot}{\cdot}$. Figure~\ref{fig:representations}
  shows how the reported complete ranking can be thus encoded: $\aD$
  is ranked over $\aB$ and $\aC$ both, therefore the corresponding
  elements in the matrix
  $\yvpref{\aA}{\aD}{\aB}=\yvpref{\aA}{\aD}{\aC}=1$, while $\aC$ is not
  ranked over $\aD$, so $\yvpref{\aA}{\aD}{\aC}=0$. Because $\aA$ does
  not rank self, the row and column corresponding to $\aA$ are
  undefined, and because it is not meaningful to compare an alter to
  itself, so is the diagonal. (Notably, if $\aA$ were allowed to rank
  self, $\aA$'s row and column would be defined, and if $\aA$ had then
  \emph{chosen} to not rank self, they would be set to $0$.) The
  collection of reported rankings by all the egos in $\actors$ can
  then then be combined into a binary
  $\nactors\times\nactors\times\nactors$-array.}

\newtxt{As we have noted, because the framework itself requires relatively
  few assumptions regarding $\pref[i]$, pairwise comparison matrices can
  encode a wider variety of ranking structures: for example, in
  $\yexw$ in Figure~\ref{fig:toy-weird}, Ego $\aC$ does not provide enough
  information to establish a weak order, while Ego $\aD$'s reports
  violate transitivity of comparisons. This precludes their representation in
  the corresponding rows of $\yexw_{\cdot,\cdot}$ but not their
  representation as $\egopref{\yexw}{\aC}{\cdot}{\cdot}$ and
  $\egopref{\yexw}{\aD}{\cdot}{\cdot}$}

\subsection{Model Formulation and Specification for ERGMs for Compete Rankings}
\newtxt{\citet{krivitsky2012erg} suggests that a sample space of complete
rankings of every actor in a network by every other actor can be
represented by a directed network with no self-loops, whose set of observed relations $\dysY=\distuples{2}$ maps to dyad values
$\dyadvals=\fromthru{1}{n-1}$, with the ranked nature of the data
leading to a complex constraint: that an ego $i$ \emph{must} assign a
unique rank to each possible alter. Formally,
\begin{equation}\netsY=\{\yv'\in \dyadvals^\dysY: \forall_{i\in N} \forall_{r\in \dyadvals}\exists!_{j\in N\setsub \{i\}} \yij'=r \}.\label{eq:sample-space}\end{equation}}

\newtxt{Again, this representation is slightly misleading in that elements of
$\dyadvals$ have only ordinal and not interval or ratio meanings, and,
as we noted above, are only ordered within the rankings of a given
ego. That is, it makes sense to ask if for some $\ynetsY$ and
$(i,j,k)\in \distuples{3}$, $\yij>\y_{i,k}$---i.e., whether $i$ ranks
$j$ over $k$---but not to evaluate the difference between ranks
($\yij-\y_{i,k}$) or to compare ranks by different egos
($\y_{j,i}>\y_{k,i}$). It does, however, represent distinct complete
rankings in a concise and convenient manner, so we make use of it,
with the proviso that \emph{the statistics evaluated on $\ynetsY$ make use
of no operation other than comparison within an ego's rankings $\ypref{i}{j}{k}$.}}

Taking the set defined by \eqref{eq:sample-space} as our sample space, we can specify an
exponential family for rank-order networks by defining a sufficient
statistic $\genstats(\yv;\xv)$, a function of a network $\ynetsY$ that may also depend on exogenous covariates $\xv\in\xs$ (assumed fixed and known), for an exponential family and parametrized by $\paramv\in\reals^\npar$. The probability associated with each network $\yv$ in the sample space is then
\begin{equation}
  \Ptegx(\Yy)=\frac{\exp\en\{\}{\innerprod{\paramv}{\genstats(\yv;\xv)}}}{\cegx(\paramv)},~\ynetsY\label{eq:rankergm}
\end{equation}
 with the normalizing factor
\begin{equation}\cegx(\paramv)=\sum_\ypnetsY \exp\en\{\}{\innerprod{\paramv}{\genstats(\yv';\xv)}}. \label{eq:rankergmc}
\end{equation}
This is an exact parallel to the more familiar ERGMs for dichotomous data \citep[e.g.,][]{wasserman1996lml}. For notational convenience, we will drop $\xv$ from now on, unless the term in question uses it explicitly.

\section{\label{sec:terms}Terms and Parameters for Ordinal Relational Data}
We now introduce and discuss a variety of sufficient statistics
$\genstats(\cdot;\cdot)$ for the model of \eqref{eq:rankergm} (\enquote{model terms})
that abide by the restrictions discussed in
Section~\ref{sec:framework} while viably representing phenomena
frequently observed in social networks. For each term, we discuss how
it may be interpreted, without assuming rankings to have more than ordinal meaning.

\subsection{\label{sec:promotion}Interpreting Model Terms Using \enquote{Promotion} Statistics}
For binary ERGMs, \citet{snijders2006nse}, \citet{hunter2008epf}, and
others have used \emph{change statistics} or \emph{change scores}, the
effect on the model of a toggling of a tie---an atomic change in the
binary network structure---to aid in interpreting the model terms. For
complete ordering data, where changing the ranking of one individual
necessarily changes ranking of others, a candidate for an analogous
atomic change must affect positions of as few actors as possible. Such
a change is the swapping of rankings by two adjacently-ranked alters
by a single ego. We thus use the effect of having ego $i$
\enquote{promote} a promotable alter $j\in\{k\in\actors:k\ne i \land
\y_{i,k}<n-1\}$, swapping $j$'s rank with that of the alter previously
ranked immediately above $j$, as such a change. \newtxt{When it is
  clear from context (as below), we will use \enquote{$\jplus$} to
  refer to the promoted-over alter.} (See \citet{butts2007pmr} for a related use of
pairwise permutations to assess model terms.)

\newtxt{Let $\egoswapr{\yv}{i}{j}{\jplus}$ represent the network $\yv$ with $i$'s ranking of $j$ and $\jplus$, the actor previously ranked by $i$ immediately above $j$, swapped.  We define a \emph{promotion statistic} as
\[\ipromotejv \genstats(\yv)\equiv \genstats(\egoswapr{\yv}{i}{j}{\jplus})-\genstats(\yv),\]
i.e., the change in $\genstats$ resulting from \enquote{promoting} $j$ by one rank (and demoting the alter above \xem). An example of this is given Figure~\ref{fig:toy}: starting with a network $\yv$ \subref{fig:toy-orig}, Ego~$\aA$ promotes Alter~$\aB$ (who had been ranked immediately below $\aC$ in $\yv$) over $\aC$ to create $\yexbc$ \subref{fig:toy-ABC}. Its effect on $\yv_{\cdot,\cdot}$ is to swap the rank values
    $\y_{\aA,\aB}$ and $\y_{\aA,\aC}$, and the effect on the pairwise comparison matrix $\yvpref{\aA}{\cdot}{\cdot}$ is the smallest possible subject to the complete ordering constraint: the indicators $\ypref{\aA}{\aB}{\aC}$ and $\ypref{\aA}{\aC}{\aB}$ are swapped. In contrast, if Ego~$\aA$ promotes Alter~$\aB$ over $\aD$, who had not been adjacently ranked to $\aB$ in $\yv$, producing $\yexbd$ \subref{fig:toy-ABD}, the effect on $\yv_{\cdot,\cdot}$ is similar as before, but the effect on $\yvpref{\aA}{\cdot}{\cdot}$ is profound.
  }

\newtxt{Analogously to change scores, the promotion statistic emerges when considering
the conditional probability of an ego $i$ ranking an alter $j$ over
alter $\jplus$, other rankings being the same. To see how, let $\yv$ be a network that has $i$ ranking $k\equiv\jplus$ immediately over $j$, so that $\egoswapr{\yv}{i}{j}{k}$ is an otherwise identical network where $i$ ranks $j$ over $k$ instead. Then,
\begin{align*}
  \Pteg(\egopref{\y}{i}{j}{k}=1|&\Yv=\egoswapr{\yv}{i}{j}{k}\lor\Yy)\\
&=\frac{\Pteg\en\{\}{\egopref{\Y}{i}{j}{k}=1\land(\Yv=\egoswapr{\yv}{i}{j}{k}\lor\Yy)}}{\Pteg(\Yv=\egoswapr{\yv}{i}{j}{k}\lor\Yy)}\\
&=\frac{\Pteg(\Yv=\egoswapr{\yv}{i}{j}{k})}{\Pteg(\Yv=\egoswapr{\yv}{i}{j}{k})+\Pteg(\Yy)}\\
&=\frac{\exp\en\{\}{\innerprod{\paramv}{\genstats\en(){\egoswapr{\yv}{i}{j}{k}}}}/\cancel{\ceg(\paramv)}} {\exp\en\{\}{\innerprod{\paramv}{\genstats\en(){\egoswapr{\yv}{i}{j}{k}}}}/\cancel{\ceg(\paramv)}+\exp\en\{\}{\innerprod{\paramv}{\genstats(\yv)}}/\cancel{\ceg(\paramv)}}\\
&=\frac{\exp\en(){\innerprod{\paramv}{\en\{\}{\genstats\en(){\egoswapr{\yv}{i}{j}{k}}-\genstats(\yv)}}}} {\exp\EN[]{\innerprod{\paramv}{\EN\{\}{\genstats\en(){\egoswapr{\yv}{i}{j}{k}}-\genstats(\yv)}}}+1}\\
&=\ilogit\en\{\}{\innerprod{\paramv}{\ipromotejv \genstats(\yv)}},
\end{align*}}
since, by construction, $k\equiv\jplus$ in $\yv$.

In this, the promotion statistic also reflects the conditional dependence
structure of the model: if its form for a particular
$\genstats(\cdot;\cdot)$ does not depend on a particular datum, using
it in the model cannot induce conditional dependence on that
datum. Thus, although we do not derive our model terms from a
conditional dependence structure (cf. \citet{robins1999lml}), we can
use them to examine the conditional dependence structure of the model
for each term we consider.

Note that promotion
statistics are mainly useful for complete orderings: if the
ordering is partial, it is possible for $i$ to promote $j$ without
demoting $\jplus$.  (See \citet{butts:sm:2007b} for a parallel case involving models for one-to-one versus many-to-one assignments.)

\subsection{\label{sec:exogenous}Terms for Exogenous Covariates}

We begin our quorum of substantively useful statistics by considering
\enquote{exogenous} factors: those factors that would influence
rankings by an ego $i$ in a manner that is independent (in the
probabilistic sense) of the rankings of all other egos
$i'\in\actorsnot{i}$. (Indeed, none of the promotion
statistics in this section depend on any other rankings.)
Substantively, these factors are exogenous to the ranking process in
that they are not, at least on the time scale of the process, mutable,
or in that they operate independently of an ego $i$ being able to
observe or infer rankings or other salient states or actions (that are
endogenous to the model) of any other ego $i'$.

\subsubsection{\label{sec:attractiveness}Attractiveness/Popularity Effects}

For assessments of attractiveness, liking, and status, it is likely
that egos' rankings will be influenced by some relatively stable (and
exogenous) tendencies of particular alters to be rated more highly than
others.  For instance, assessments of physical attractiveness tend to
be broadly consistent within a given cultural context, and such
assessments correlate positively with physical attributes and
performance characteristics (e.g., subtleties of dress and speech)
that are usually difficult to alter over short time scales \citep{morse.et.al:jp:1974,webster.driskell:ajs:1983}.
Thus, we have the emic notion that some persons \enquote{are
  attractive}, with the attribution regarded as a fixed trait of the
person being assessed; while the reality is less trivial, stable
factors governing attractiveness are sufficiently important that we
may wish to capture them where possible.  In other settings,
institutionalized status characteristics (e.g., group membership,
formal social roles) or the like may have similar effects \citep{berger.et.al:asr:1972,berger.et.al:bk:1977}.

Regardless of source, we can treat these effects directly by positing some covariate vector $\xv\in \reals^{\nactors}$, associated with a statistic of the general form
\[\genstat{A}(\yv;\xv) = \sum_{(i,j,k)\in \distuples{3}} \ypref{i}{j}{k}(\x_j-\x_k).\]
This statistic simply indexes the tendency for
those with higher values on $\x$ to be ranked more highly than
those with lower values.  The promotion statistic associated with the above is of the form
\[
  \ipromotej \genstat{A}(\yv;\xv)= 2(\x_{j}-\x_{\jplus}),
\]
i.e., twice the difference between the attractiveness of $j$ and the actor
over whom $j$ may be promoted.  Thus, the effect of this term on the
odds of $i$ promoting $j$ is
$\myexp{2\param_{\text{A}}(\x_{j}-\x_{\jplus})}$.
As expressed, $\genstat{A}$ treats $\xv$ as at least an
interval scale; in other cases, the subtraction operator would need to
be replaced with a more appropriate function.  While $\xv$ may
be an observed covariate, it is worth noting that this quantity is
also a natural candidate for treatment via a random popularity effect \citep{duijn2004re}.

\paragraph{Example:}

\newtxt{As an illustrative example of the above, we consider these statistics in the cases of the sample complete-ordering networks $\yex$ and $\yexbc$ from Figure~\ref{fig:toy}.  For purposes of illustration, we also assume that each sample network is associated with a covariate vector $\xv$ taking the values $(1,2,3,4)$.  Note that $\genstat{A}$ is equal to the sum of the signed differences in the elements of $x$ for all those ordered pairs having non-zero entries in $\ypref{\cdot}{\cdot}{\cdot}$.  These entries are depicted in Figure~\ref{fig:toy}; summing the associated differences gives us $\genstat{A}(\yex;\xv)=6$ and $\genstat{A}(\yexbc;\xv)=4$.  Intuitively, the ratings of $\yex$ are more strongly aligned with the attribute values in $\xv$ than those of $\yexbc$.  Note that the difference between $\genstat{A}(\yex;\xv)$ and $\genstat{A}(\yexbc;\xv)$ is necessarily equal to the promotion statistic for a swap of $\aB$ and $\aC$ in $\aA$'s rating structure, i.e. $2(\x_{\aB}-\x_{\aC})=2(2-3)=-2$.  In the case of attractiveness, we thus have the intuitive notion that promoting an actor with a lower attribute value over one whose attribute value is higher results in a decline in the associated statistic.}

\subsubsection{Difference/Similarity Effects}

Just as one may posit a differential tendency to \enquote{win} ranking
contests overall, one may also posit that each actor $i$ has exogenous
characteristic $\xv_i\in\xs$ such that alters
\enquote{close to} or \enquote{far from} ego will be more likely to be
highly ranked than those with the reverse attributes.  (Such assumptions are the basis of models such as e.g., spatial voting theory \citep{enelow.hinich:bk:1984}.)  This is a
familiar application of homophily/heterophily to the rank order case,
and the implementation is straightforward:
\begin{equation}
\genstat{H}(\yv;\xv) = \sum_{(i,j,k)\in \distuples{3}} \ypref{i}{j}{k}\left[z(\xv_i,\xv_j)-z(\xv_i,\xv_k)\right], \label{eq:homophily}
\end{equation}
where $z:\xs^2\to \reals$ is any function that is monotone
increasing in the difference between its arguments.  Thus, where this
statistic is enhanced we expect \enquote{far} actors to outrank
\enquote{near} ones (from the point of actor $i$), with the reverse
holding where this statistic is suppressed. The atomic effect of this
term is simply
\[
  \ipromotej \genstat{H}(\yv;\xv)= 2\left[z(\xv_{i},\xv_{j})-z(\xv_{i},\xv_{\jplus})\right].
\]
As with attractiveness, difference effects can be based either on observed covariates or on latent quantities.

\newtxt{\paragraph{Example:} Returning}
\newtxt{to our two sample cases, let us consider how $\genstat{H}$ varies (again taking $\xv=\{1,2,3,4\}$, and for simplicity taking $z$ equal to the absolute difference).  In the case of $\genstat{A}$, our statistic reduced to the sum of attribute value differences for ordered pairs having a judgment in $\ypref{\cdot}{\cdot}{\cdot}$; the present case can be recognized as essentially similar, with the only distinction being that the relevant attribute for $\ypref{i}{j}{k}$ is now the proximity to the rater ($i$) vis-a-vis $\xv$.  Summing over the ordinal judgments in each case gives us $\genstat{H}(\yex;\xv)=6$, $\genstat{H}(\yexbc;\xv)=4$, and $\promote{\aA}{\aB}\genstat{H}(\yex)=2(\abs{\x_{\aA}-\x_{\aB}}-\abs{\x_{\aA}-\x_{\aC}})=2(\abs{1-2}-\abs{1-3})=-2$.}

\subsubsection{Dyadic Covariates} \label{sec:dyadic}

We can extend the above logic to general dyadic covariates.  For
instance, we may consider a case in which a within-context ranking is
made by actors having ongoing social relationships; we might expect,
then, that actors engaged in positive long-term relationships would
tend to give preference to their partners within the specific rating
context.  Statistics for this behavior can be produced like so:
\[
\genstat{Dyad}(\yv;\xv) = \sum_{(i,j,k)\in \distuples{3}} \ypref{i}{j}{k}\left(\x\sij-\x\sik\right)
\]
\[
  \ipromotej \genstat{Dyad}(\yv;\xv)= 2\left(\x\sij-\x_{i,\jplus}\right).
\]
Of course, the cases of attractiveness and difference described above
are simply special cases of dyadic covariates, with particular
structure imposed. (Notably, the matrix permutation family (ERGP)
of \citet{butts2007pmr} has a somewhat similar structure.)  As such, we do not provide numerical examples here (though see the empirical examples in section~\ref{sec:bk}.)

\subsubsection{Comparison Covariates}

Finally, in the framework of pairwise comparison, the most general
exogenous covariate form assigns a weight to each pairwise comparison
by each ego:
\[
\genstat{P}(\yv;\xv) = \sum_{(i,j,k)\in \distuples{3}} \ypref{i}{j}{k}\x\sijk,
\]
for some $\xv\in\reals^{\distuples{3}}$---assigning to each distinct ordered triple $(i,j,k)$ a covariate value $\x_{i,j,k}$---resulting in
\[
  \ipromotej \genstat{P}(\yv;\xv)= \left(\x_{i,j,\jplus}-\x_{i,\jplus,j}\right).
\]
This statistic has all other exogenous statistics as special cases.

\subsection{Terms for Endogenous Mechanisms}

We now turn to factors that are endogenous in the sense that, unlike
exogenous factors, their effect on the rankings by ego $i$ does depend
on rankings by other egos $i'\in\actorsnot{i}$. Substantively,
these are factors for phenomena that may plausibly arise in cases for which ego observes or is able to infer the rankings of others.

\subsubsection{\label{sec:conformity}Global Conformity}

In many settings where an ego is able to observe or infer the rankings of others,
there is reason to presume that this will influence ego, so that \xe{}
will tend to bring \xer{} own rankings into conformance with
others'.  This is certainly true in dominance or status rankings,
where there is considerable evidence that individuals can and do infer
status ordering from observation of third-party judgments \citep[see, e.g.,][]{anderson.et.al:jpsp:2006}; 
this synchronization may even be explicit, as in certain types of
gossip (wherein two or more parties \enquote{compare notes} on the
relative status of their peers) \citep{dunbar:bk:1997}.  The status of influence for
relations such as relative liking is less clear, but still plausible:
ego may take alter's evaluations of the relative merits of other
alters into account in assessing \xer{} own preferences, just one
can be influenced in one's judgment of the merits of food, art, or
other experiential goods by the evaluations of others \citep{bordieu:issj:1968}.  Finally, the
mutual observability of rankings may produce in some settings a form
of \enquote{conformity pressure} \citep[, for example]{asch1951egp}, such that those
displaying deviant rankings anticipate (and are possibly exposed to)
sanction.  The importance of influence processes in such settings is
well-documented.

To formalize influence in the ranking context, we must note that four
elements are involved: ego's assessment of two alters (say, $j$ and
$k$), and the assessment of those same alters by a distinct third
party (say, $\l \notin\{i,j,k\}$).  Denoting ego by $i$, we note that
ego's assessment of $j$ and $k$ is in conformity with $\l$'s
assessment of $j$ and $k$ when
$\ypref{i}{j}{k}=\ypref{\l}{j}{k}$ and
$\ypref{i}{k}{j}=\ypref{\l}{k}{j}$.  A natural statistic
to summarize the degree of ratings nonconformity, then, is
\newtxt{\begin{equation}
\genstat{GNC}(\yv) = \sum_{(i,j,k,\l)\in \distuples{4}}\ypref{\l}{j}{k}\left(1-\ypref{i}{j}{k}\right). \label{eq:nonconformity}
\end{equation}}
The promotion statistic for nonconformity can be derived by observing that when $i$ promotes $j$ over $\jplus$, the statistic is incremented by 2 every other ego $\l$ who has $\jplus$ ranked over $j$ and decremented by 2 for every $\l$ who has $j$ ranked over $\jplus$ \newtxt{(1 for $i$ conforming/disconforming to $\l$ and 1 for $\l$ conforming/disconforming to $i$)}. Thus,
\newtxt{\begin{equation}
  \ipromotej \genstat{GNC}(\yv) = 2\sum_{\l \in \actorsnot{i,j,\jplus}}\en(){\ypref{\l}{\jplus}{j}-\ypref{\l}{j}{\jplus}}, \label{eq:nonconformity-promote}
\end{equation}}
a \enquote{vote} among the $\l$s as to the relative ranking of $j$ and $\jplus$. Insofar as influence is active, $\genstat{GNC}$ should be suppressed (and
the associated parameter negative).  In the typical case of total
ordering within subjects, sufficiently strong suppression of
$\genstat{GNC}$ will force the population to converge to a universal
consensus ranking; if the suppression is weaker, a looser but
analogous set of states will be favored.

It should be noted that this form of influence and the random attractiveness effect, mentioned in
Section~\ref{sec:attractiveness}, can both explain the same network feature: both heterogeneity in attractiveness
and social influence induce an agreement in rankings, and the
latter may be considered a marginal representation of the former, in a manner similar to that of a within-group
correlation as a marginal reflection of a (conditional) random effects
linear model.

\newtxt{\paragraph{Example:} For our sample networks, $\yex$ and $\yexbc$, summing over all tetrads gives us respectively $\genstat{A}(\yex)=8$ and $\genstat{A}(\yexbc)=6$.  As this implies, $\promote{\aA}{\aB}\genstat{GNC}(\yex)=2\sum_{\l \in \actorsnot{\aA,\aB,\aC}} \en(){\ypref{\l}{\aC}{\aB}-\ypref{\l}{\aB}{\aC}}=2 \en(){\ypref{\aD}{\aC}{\aB}- \ypref{\aD}{\aB}{\aC}}=2(0-1)=-2$.  Intuitively, this implies that the ratings of $\yexbc$ are in greater consistency with one another than those of $\yex$; as the promotion statistic indicates, this is because $\aA$'s switching from $\aC\pref[\aA]\aB$ to $\aB\pref[\aA]\aC$ brings $\aA$ view of $\aB$ and $\aC$ into conformity with the view of $\aD$.  To the extent that greater conformity pressure is present (i.e., a negative coefficient associated with $\genstat{A}$, such changes are (\emph{ceteris paribus}) favored.}

\subsubsection{Local Conformity}

For global nonconformity $\genstat{GNC}$, the promotion statistic~\eqref{eq:nonconformity-promote} implies that $i$ weights agreement
with every other $\l$ equally, regardless of how $i$ had ranked $\l$.  In some cases, it may be plausible that the salience of $\l$ for $i$ may depend upon $i$'s ranking of \xem; for instance, $i$ may be more likely to attend to (and to conform to) those whom \xe{} ranks highly
than those whom \xe{} ranks lower. A possible formalization of this is
the notion that $i$'s ranking of $j$ would be influenced by $\l$ only
if $i$ ranks $\l$ over $j$, so only actors ranked above $j$ influence
$i$'s rankings involving $j$. As with global conformity, we define this
statistic, the \emph{local nonconformity}, negatively: counting the
number of instances where an ego had ranked $\l$ \newtxt{over two alters $j$ and $k$ but then did
not conform to $\l$'s ranking of $j$ relative to $k$}:
\begin{equation}
\genstat{LNC}(\yv) = \sum_{(i,j,k,\l)\in \distuples{4}} \ypref{i}{\l}{j} \newtxt{\ypref{i}{\l}{k}} \ypref{\l}{j}{k} (1-\ypref{i}{j}{k}). \label{eq:local-nonconformity}
\end{equation}
The atomic effects for this statistic are somewhat complex:
\newtxt{\begin{subequations}
\begin{align}
  \ipromotej \genstat{LNC}(\yv)=\sum_{k\in \actorsnot{i,j,\jplus}}(&  \ypref{i}{k}{\jplus}\ypref{k}{\jplus}{j}-\ypref{i}{k}{\jplus}\ypref{k}{j}{\jplus}\label{eq:lnc-conforms}\\
  \vphantom{\sum_{k\in \actorsnot{i,j,\jplus}}}&+\ypref{k}{i}{\jplus}\ypref{k}{\jplus}{j}-\ypref{k}{i}{j}\ypref{k}{j}{\jplus}\label{eq:lnc-leads}\\
  \vphantom{\sum_{k\in \actorsnot{i,j,\jplus}}}&+\ypref{j}{k}{\jplus}\ypref{i}{\jplus}{k}-\ypref{\jplus}{k}{j}\ypref{i}{j}{k}). \label{eq:lnc-discounts}
\end{align}
\end{subequations}}
They have, however, a meaningful interpretation. The two terms \eqref{eq:lnc-conforms} represent the effect of $i$ bringing \xer{} ordering of
$j$ and $\jplus$ into conformance (or disconformance) with those of
some actor $k$ whom $i$ had ranked higher than at least one of
them. The pair \eqref{eq:lnc-leads} represents the situation where some actor $k$
ranks $i$ over $j$ and/or $\jplus$, so $i$ promoting $j$ over $\jplus$
either creates or eliminates disconformance on the part of $k$. The
pair \eqref{eq:lnc-discounts} represent the notion that nonconformity is also
created if actors $i$ and $j$ disagree on the ordering of $\jplus$ and
some actor $k$, so $i$ promoting $j$ over $\jplus$ creates
disconformance by making $j$'s ordering of $\jplus$ salient to
$i$. (Ego $i$ can resolve this tension either by changing the ranking of
$\jplus$ and $k$ to conform with $j$ (affecting \eqref{eq:lnc-conforms})
or by demoting $j$ to make \xer{} ranking less salient.)

\newtxt{\paragraph{Example:}  For our sample networks $\yex$ and $\yexbc$, calculation of the local conformity statistic shows that $\yex$ exhibits a slightly higher level of non-conformity than $\yexbc$: $\genstat{LNC}(\yex)=2$, versus $\genstat{LNC}(\yexbc)=1$.  To appreciate the origin of this difference we may examine the promotion statistic:
\begin{align*}
\promote{\aA}{\aB}\genstat{LNC}(\yex)=(&  \ypref{\aA}{\aD}{\aC}\ypref{\aD}{\aC}{\aB}-\ypref{\aA}{\aD}{\aC}\ypref{\aD}{\aB}{\aC}\\
  &+\ypref{\aD}{\aA}{\aC}\ypref{\aD}{\aC}{\aB}-\ypref{\aD}{\aA}{\aB}\ypref{\aD}{\aB}{\aC}\\
  &+\ypref{\aA}{\aC}{\aD}\ypref{\aB}{\aD}{\aC}-\ypref{\aA}{\aB}{\aD}\ypref{\aC}{\aB}{\aD})\\
=(&1\times 0-1\times 1\\
  &+0\times 0-0\times 1\\
  &+0\times 1-0\times 1)=-1.
\end{align*}
In promoting $\aB$ over $\aC$, $\aA$ conformed with $\aD$---the only actor $\aA$ ranked over both $\aB$ and $\aC$---so local nonconformity declined (per Equation \eqref{eq:lnc-conforms}).  $\aD$ did not rank $\aA$ highly enough to experience conformity pressure in ranking $\aB$ and $\aC$, and the only actor who ranked $\aA$ highly is $\aC$, but since actors are not permitted to rank themselves, $\aC$ has nothing to say about C's own ranking relative to $\aB$ (see Equation~\eqref{eq:lnc-leads}).  Promoting $\aC$ over $\aB$ made $\aB$ salient in comparisons about $\aC$, but not about $\aD$, so no disconformance was created or eliminated in this respect (see Equation~\eqref{eq:lnc-discounts}).  As this case illustrates, local nonconformity involves both consistency of ratings and status: to the extent that pressure for local conformity is present, actors will seek to adapt their rankings to match those of individuals they rate highly, with less regard to mismatches with those to whom they assign a low rank.
}

\subsubsection{Deference Aversion}

Influence, as defined above, deals with the mutual adjustment among
raters regarding their relative assessments of third parties.  When
ego is a party to the rating in question, the situation becomes more
complex.  By assumption, ego does not explicitly self-rate; thus, ego
cannot adjust towards alter's impression of \xem.  In many
settings, however, another mechanism may be active that will make
alter's ranking of ego salient for ego's ranking of alter.  In
particular, consider the case in which higher rankings are associated
with positive evaluation, such that being ranked below others is
aversive.  Moreover, let us assume that ego infers \xer{} own
status via an \emph{implicit transitivity mechanism}, such that if
alter $\l$ ranks $j$ above ego ($i$) and ego ranks $\l$ above $j$,
then ego is for social purposes ranking \xemself{} below $\l$.
Under such circumstances, \emph{deference aversion} may lead ego to
resist ranking $\l$ above $j$.

To capture this notion with a statistic, we propose the following:
\begin{equation}
\genstat{D}(\yv) = \sum_{(i,j,\l)\in \distuples{3}} \ypref{\l}{j}{i}\ypref{i}{\l}{j}. \label{eq:deference}
\end{equation}
We expect this statistic to be suppressed where deference aversion is
present. The promotion statistic is incremented if $\jplus$ had ranked $i$ over $j$,
since it creates a deference of $\jplus$ to $i$ via $j$; or if $j$ had
ranked $\jplus$ over $i$ since it creates a deference of $i$ to $j$ via
$k$; and it is decremented if $\jplus$ had ranked $j$ over $i$, as it would
eliminate the deference of $i$ to $\jplus$ via $j$; or if $j$ had ranked
$i$ over $\jplus$, as it would eliminate the deference of $j$ to $i$ via
$\jplus$. Thus,
\begin{align*}
  \ipromotej \genstat{D}(\yv) = 2\en(){\ypref{\jplus}{i}{j}+\ypref{j}{\jplus}{i} - 1}.
\end{align*}
It is interesting to note that the principal effect of suppressing
this statistic is actually to bring ego's rankings in line with those
of alter, somewhat akin to the reciprocity or mutuality in binary
relations.  Specifically, if there are $r$ persons ranked by alter as
being above ego, then ego will also tend to rank those same $r$
persons as being above alter.  Where a total order is present, ego and
alter will thus tend to give each other the same rank (and, indeed, to
agree on those persons having higher ranks).  Of course, applying this
logic to all pairs suggests pressure towards equality, which is
impossible to achieve in the total order case (but not necessarily
for others).  Even in the case of total orders, however, considerable
variation in $\genstat{D}$ is possible, with lower values indicating
rating structures in which agreement between raters on high-ranked
alters is maximized.

\newtxt{\paragraph{Example:} While we saw that $\yexbc$ showed lower local nonconformity than $\yex$, we see the opposite for deference aversion: applying the above definition to the two networks gives us $\genstat{D}(\yex)=6$ and $\genstat{D}(\yexbc)=8$, implying that $\yex$ has fewer instances in which an individual tacitly places him or herself below others in the network.  Examining the promotion statistic, we see that
\begin{align*}
\promote{\aA}{\aB} \genstat{D}(\yv)& = 2\en(){\ypref{\aC}{\aA}{\aB}+\ypref{\aB}{\aC}{\aA} - 1}\\
& = 2\times( 1 +  1 - 1) =2.
\end{align*}
Intuitively, $\aC$ had ranked $\aA$ over $\aB$, so in ranking $\aC$ over $\aB$, $\aA$ was not deferring to either party.  By contrast, $\aB$ had ranked $\aC$ over $\aA$, so in promoting $\aB$ over $\aC$, $\aA$ implicitly deferred to $\aB$.  In settings for which deference aversion is operative, such changes are \emph{ceteris paribus} unfavorable.}

\subsection{Consistency Across Settings} \label{sec:consistency}

When ranking the same alters among multiple settings---across time
or across rubrics---there is reason to expect that ego will tend to
exhibit consistency in alter ratings. Across time, this is an
exogenous effect, because earlier rankings cannot be influenced by
later rankings. Across rubrics, it may be endogenous.  Here, we assume
two rating structures, $\yv$ and $\yv'$, on vertex sets $\actors$ and
$\actors'$, such that some set $\actors_s=\actors \cap \actors'$ of
actors are involved in both networks.  For convenience in notation, we
take the labeling of the members of $\actors_s$ to be the same in both
$\actors$ and $\actors'$.  Given this, our statistic measuring
inconsistency is simply
\begin{equation}
\genstat{I}(\yv;\yv') = \sum_{(i,j,k)\in\distuples{3}_s} \left[ \ypref{i}{j}{k}\en(){1-\egopref{\y'}{i}{j}{k}} + \left(1-\ypref{i}{j}{k}\right) \egopref{\y'}{i}{j}{k} \right], \label{eq:inconsistency}
\end{equation}
with promotion statistic being simply 
\begin{align*}
\ipromotej \genstat{I}(\yv;\yv') = 2(\egopref{\y'}{i}{\jplus}{j}-\egopref{\y'}{i}{j}{\jplus}).
\end{align*}
As $\genstat{I}$ measures the discordant pairs of rankings in $\yv$
versus $\yv'$, suppressing it implies higher levels of cross-context
consistency. 

The statistic \eqref{eq:inconsistency} treats all disagreements between $\yv$ and $\yv'$ as equivalent. It may be the case, however, that only some disagreements are of interest, or disagreements themselves need to be modeled. This can be facilitated by a more general form of $\genstat{I}$. Given weights $\xv \in \reals^{\distuples{3}}$, (symmetric for complete orderings, such that $\x\sijk\equiv\x_{i,k,j}$), let the \emph{weighted inconsistency} of $\yv$ versus $\yv'$ be defined as
\begin{equation}
\genstat{I}(\yv;\yv',\xv) = \sum_{(i,j,k)\in\distuples{3}_s} \left[ \ypref{i}{j}{k}\en(){1-\egopref{\y'}{i}{j}{k}} + \left(1-\ypref{i}{j}{k}\right) \egopref{\y'}{i}{j}{k} \right]\x\sijk. \label{eq:wt-inconsistency}
\end{equation}
Note that $\xv$ can, itself, be parametrized to model factors affecting
disagreement between two rankings, making it a potentially interesting basis for hierarchical modeling.

\newtxt{These statistics are utilized extensively in the example of Section~\ref{sec:bk}, where they are employed to examine the accuracy of informants' self-reported interaction frequencies.}

\section{\label{sec:inference}Inference and Implementation}
The proposed class of models is a finite exponential family over a
finite sample space. Its natural parameter space
$\natcurvpars=\{\paramv':\ceg(\paramv')<\infty\}=\reals^\npar$, an
open set, so the models from this family are regular and posses all of
the inferential properties of such
families \citep[:1--2]{brown1986fse}.  Here, we briefly discuss practical issues associated with fitting the models
of this class.

\subsection{Implementation}

The sample space of a complete ranking ERGM is finite but large
($\abs{\netsY}=\en\{\}{(\nactors-1)!}^\nactors$), so even for networks
of modest size, evaluating the normalizing constant
\eqref{eq:rankergmc} is not computationally feasible. However,
provided a method for sampling from this ERGM distribution
$\Pteg(\cdot)$ exists, the Monte Carlo MLE method can be used to fit
the model. \citep{geyer1992cmc,hunter2006ice,krivitsky2012erg} We
\newtxt{review this technique and} describe a simple algorithm for
sampling from an ERGM for complete ordering defined in
Section~\ref{sec:framework} in the Appendix. \newtxt{Bayesian
  inference using exchange algorithms is also possible, if more
  computationally expensive. \citep{CaFr11b}}

We base our implementation on the valued ERGM extensions of
\citet{krivitsky2012erg} to the \proglang{R} \citep{RC14r} package \pkg{ergm} \citep{hunter2008epf,handcock2010epf}. We expect to publicly
release our extensions in a new package, \pkg{ergm.rank}.

Lacking a general procedure such as the maximum pseudo-likelihood
estimation (MPLE) \citep{strauss1990pes} to use as a starting point
for optimization, we initiated the MCMC MLE optimization at
$\paramv=\0$ in each of our applications. This parameter configuration
corresponds to a uniform distribution on the space of possible
rankings\newtxt{, and it is therefore the safest
  configuration to start with}. Although a poor starting value can
cause MCMC MLE to fail, techniques such as the Stepping Algorithm of
\citep{HuHu12i} can be used to ameliorate its effects, and a form of
it is used here as well.

\subsection{Degeneracy}

Degeneracy, in its many meanings, is
often a concern in applying ERGMs. Despite recent progress \citep[, for example]{rinaldo2009gde,butts2011bgb,schweinberger2011isd}, few general and unambiguous diagnostics for degeneracy
exist beyond simulated goodness-of-fit measures \citep{hunter2008gfs}
and diagnostics of symptoms such as poor convergence of the
MLE-finding procedure and, in particular, poor mixing of the MCMC MLE
algorithm \citep{goodreau2008st}, often due to multimodality of the
distribution of networks under the model \citep{snijders2006nse,rinaldo2009gde,krivitsky2012erg}.

In our case, there is some cause for concern due to the high order of
some of the sufficient statistics we propose. For example,
$\genstat{GNC}$ is a summation of indicators over a set of
cardinality $\abs{\distuples{4}}=O(\nactors^4)$. Its promotion statistic
$\ipromotej \genstat{GNC}$ is a sum over a set of $O(\nactors)$. Thus,
there may exist configurations for which a small change in the ranking
leads to a massive change in the value of this statistic and thus in
the likelihood. Such statistics can induce excessively strong dependence in many cases lead to asymptotic degeneracy \citep{butts2011bgb,schweinberger2011isd}.

At the same time, a number of factors are likely to ameliorate such
problems for the specific case of rank data. A network of ranks simply
contains more information than a network of binary outcomes of the
same size. More concretely, a uniform distribution over binary
directed networks of size $\nactors$ with no self-loops has entropy of
$\nactors(\nactors-1)$ bits, while a uniform distribution over rank
networks has entropy of
$\log_2\en\{\}{(\nactors-1)!}^\nactors=\nactors\log_2\en\{\}{(\nactors-1)!}$
bits, which grows much faster in $\nactors$. This can reduce the risk of pathological model behavior due to poor identification in the small-$\nactors$ case.  Also, while there exists
the potential for problematic statistics and configurations, the
constrained nature of the sample space makes degeneracy more difficult to achieve than is the case for typical graphs or digraphs: an ego
cannot promote one alter without demoting another, and while some
statistics, like $\genstat{GNC}$, have an extreme case of all actors
agreeing, others, like $\genstat{D}$, cannot be reduced to 0.

In all of the following examples, we found no symptoms of degeneracy:
all estimators converged without issues to estimates for which the
expectations of sufficient statistics \newtxt{from the penultimate MCMC MLE step were close to} their observed values,
and MCMC diagnostics showed no
sign of poor mixing or multimodality, and MCMC sample sizes used were
sufficient to render the error due to the use of MCMC to approximate
the likelihood \citep[:~eq.~3.11]{hunter2006ice} negligible compared
to the standard error. That these MLEs were found and converged to
despite the $\0$ starting point, again, suggests that models using the
terms introduced here are quite robust.  In practice, we recommend
simulation from fitted models (as in the case of binary ERGMs) as a
useful tool for verifying non-degeneracy.

\section{\label{sec:examples}Examples}
\subsection{\label{sec:newcomb}Dynamics of the Acquaintance Process}
From 1953 to 1956, a research group led by Theodore
\citet{newcomb1961ap} conducted an experimental study of acquaintance
and friendship formation. In each of the two study years, 17 men
attending the University of Michigan---all transfer students with no
prior acquaintance among them---were recruited to live in off-campus
fraternity-style housing. Demographic, attitudinal, and sociometric
information was collected about the subjects. In particular, in the
second year, at each of 15 weekly time points (with week 9 being
missing), each participant was asked to rate all other participants on
\enquote{favorableness of feeling}, with ratings forced to be distinct
and converted to ranks. \citeyearpar[:32--34]{newcomb1961ap} These
data represent an example of longitudinal data of complete ranks, and,
particularly the data from the second year of the study, have been
used to study the formation of interpersonal relationships by
\citet{newcomb1956pia}, \citet{breiger1975acr}, \citet{white1976ssm},
\citet{arabie1978cbh}, \citet{wasserman1980asn},
\citet{pattison1982asm}, \citet{nakao1993las}, \citet{doreian1996bhb},
\citet{krackhardt2007hvs}, and many others.

We use ERGMs for rank-order data to study this network, examining the
social forces relevant to its structure and its evolution over
time. We take two distinct approaches: cross-sectional, where each
time point's network structure is modeled on its own and dynamic,
where each time point but the first is effectively modeled as a change
from the previous time point's rankings.

\subsubsection{Cross-sectional Analysis}

Demographic data, including age, religion, and political views of the
subjects were gathered. Also, within the house, the subjects were
assigned to rooms, spread over two floors of the house---three
one-occupant rooms, four two-occupant, and two three-occupant
rooms \citep[:67--68]{newcomb1961ap}. Furthermore, while some were
assigned to rooms at random, others were assigned with an aim to
maximize (for some rooms) and minimize (for other rooms) the
roommates' compatibility as understood by the researchers
\citeyearpar[:216--220]{newcomb1961ap}. If available, all of these factors could
be used as predictors in our modeling framework via terms introduced in Section~\ref{sec:exogenous}. Sadly, however, none of these elements of the \citeauthor{newcomb1961ap} data survive, leaving us to focus on endogenous effects (although
the \enquote{birds of a feather or friend of a friend}
\citep{goodreau2008bff} caveat applies). For each of the
15 networks, we model deference aversion (via $\genstat{D}(\Yvat{})$ \eqref{eq:deference}) and global (via $\genstat{GNC}(\Yvat{})$ \eqref{eq:nonconformity}) and local (via $\genstat{LNC}(\Yvat{})$ \eqref{eq:local-nonconformity}) conformity.
Per Section~\ref{sec:conformity}, the suppressing
of the global nonconformity statistic produces the same effect as
latent actor attractiveness, so it may, in this case, be viewed as
modeling latent heterogeneity in popularity.

\begin{figure}
  \centerline{\includegraphics[width=1\textwidth]{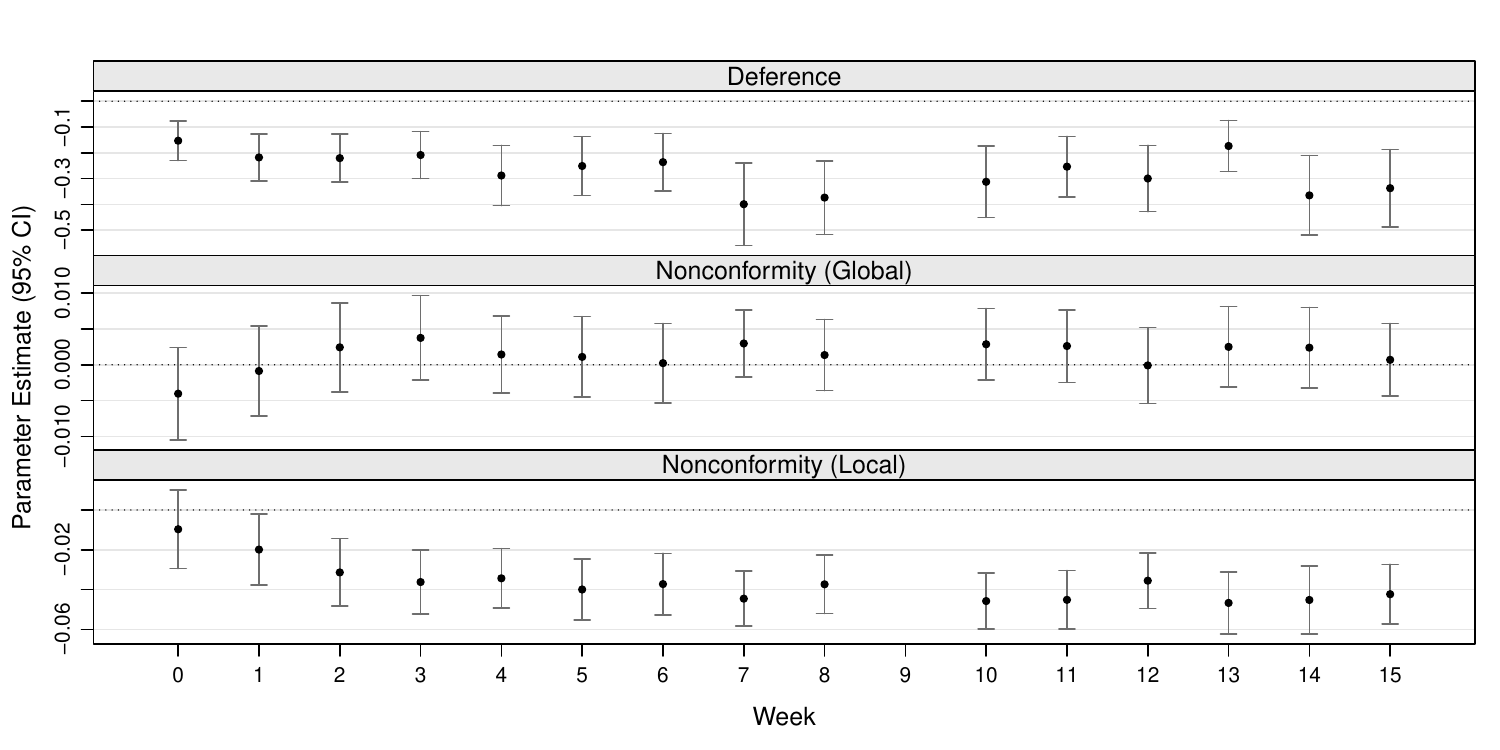}}
  \caption{\label{fig:cross-coefs}Estimated coefficients for the
    cross-sectional model fit to each week's rankings in the \citeauthor{newcomb1961ap}'s
    Fraternity. Error bars are at 95\% confidence.}
\end{figure}

\begin{table}
\begin{center}
\caption{\label{tab:cross-coefs}Results for cross-sectional analysis of the \citeauthor{newcomb1961ap}'s data }
{ \small
\begin{tabular}{rrrr}
\hline
 & \multicolumn{3}{c}{Estimates (Std. Errors)} \\
\cline{2-4}
 & & \multicolumn{2}{c}{Nonconformity} \\
\cline{3-4}
\centercol{Week} & \centercol{Deference} & \centercol{Global} & \centercol{Local} \\
\hline
0 & \coef{-0.153}{0.039}{\star\star\star} & \coef{-0.004}{0.003}{\phantom{\star}\phantom{\star}\phantom{\star}} & \coef{-0.010}{0.010}{\phantom{\star}\phantom{\star}\phantom{\star}} \\
1 & \coef{-0.218}{0.047}{\star\star\star} & \coef{-0.001}{0.003}{\phantom{\star}\phantom{\star}\phantom{\star}} & \coef{-0.020}{0.009}{\star\phantom{\star}\phantom{\star}} \\
2 & \coef{-0.221}{0.047}{\star\star\star} & \coef{0.002}{0.003}{\phantom{\star}\phantom{\star}\phantom{\star}} & \coef{-0.031}{0.009}{\star\star\star} \\
3 & \coef{-0.209}{0.046}{\star\star\star} & \coef{0.004}{0.003}{\phantom{\star}\phantom{\star}\phantom{\star}} & \coef{-0.036}{0.008}{\star\star\star} \\
4 & \coef{-0.288}{0.060}{\star\star\star} & \coef{0.001}{0.003}{\phantom{\star}\phantom{\star}\phantom{\star}} & \coef{-0.034}{0.008}{\star\star\star} \\
5 & \coef{-0.251}{0.058}{\star\star\star} & \coef{0.001}{0.003}{\phantom{\star}\phantom{\star}\phantom{\star}} & \coef{-0.040}{0.008}{\star\star\star} \\
6 & \coef{-0.236}{0.057}{\star\star\star} & \coef{0.000}{0.003}{\phantom{\star}\phantom{\star}\phantom{\star}} & \coef{-0.037}{0.008}{\star\star\star} \\
7 & \coef{-0.399}{0.081}{\star\star\star} & \coef{0.003}{0.002}{\phantom{\star}\phantom{\star}\phantom{\star}} & \coef{-0.045}{0.007}{\star\star\star} \\
8 & \coef{-0.373}{0.073}{\star\star\star} & \coef{0.001}{0.003}{\phantom{\star}\phantom{\star}\phantom{\star}} & \coef{-0.037}{0.007}{\star\star\star} \\
10 & \coef{-0.312}{0.070}{\star\star\star} & \coef{0.003}{0.003}{\phantom{\star}\phantom{\star}\phantom{\star}} & \coef{-0.046}{0.007}{\star\star\star} \\
11 & \coef{-0.254}{0.060}{\star\star\star} & \coef{0.003}{0.003}{\phantom{\star}\phantom{\star}\phantom{\star}} & \coef{-0.045}{0.008}{\star\star\star} \\
12 & \coef{-0.299}{0.066}{\star\star\star} & \coef{-0.000}{0.003}{\phantom{\star}\phantom{\star}\phantom{\star}} & \coef{-0.036}{0.007}{\star\star\star} \\
13 & \coef{-0.174}{0.050}{\star\star\star} & \coef{0.002}{0.003}{\phantom{\star}\phantom{\star}\phantom{\star}} & \coef{-0.047}{0.008}{\star\star\star} \\
14 & \coef{-0.365}{0.078}{\star\star\star} & \coef{0.002}{0.003}{\phantom{\star}\phantom{\star}\phantom{\star}} & \coef{-0.045}{0.009}{\star\star\star} \\
15 & \coef{-0.337}{0.076}{\star\star\star} & \coef{0.001}{0.003}{\phantom{\star}\phantom{\star}\phantom{\star}} & \coef{-0.042}{0.008}{\star\star\star} \\
\hline
\multicolumn{4}{l}{Significance levels: $0.05 \ge ^{\star} > 0.01 \ge ^{\star\star} > 0.001 \ge ^{\star\star\star}$}
\end{tabular}
}
\end{center}
\end{table}

We report the maximum likelihood estimates for each of the terms over
time in Table~\ref{tab:cross-coefs}
and plot them in Figure~\ref{fig:cross-coefs}.
Deference aversion is significant (the coefficient on the deference
statistic is negative) throughout the evolution of rankings, starting
with the first point of observation. This is consistent with the
finding of \citet{newcomb1956pia}, \citet{doreian1996bhb}, and others
that \enquote{friendships are reciprocated immediately}. Our analysis,
however, suggests deepening deference aversion over time, not reaching
its ultimate magnitude until week 4 or 7. (Informally, the estimated
Kendall's rank correlation between the parameter estimate and week
number is $\hat{\tau}=-0.41$, significant with
$P\text{-value}=0.036$.) One explanation for this difference is that
prior analyses, including that of \citet{doreian1996bhb}, dichotomized
the dyads in the network. Our approach uses the entire ranking and may
thus be more \newtxt{precise}.

The local nonconformity term is also significant for all but the first
observation point, although its effects seem to emerge more gradually
than those of deference aversion (Kendall's $\hat{\tau}=-0.64$,
$P\text{-value}<0.001$), leveling off around Week 7 or perhaps even
later. This does agree with \citet{doreian1996bhb}, although others
have suggested earlier and later times when the network stabilized.

In the presence of the local nonconformity term, the global nonconformity
term is not significant: there does not appear to be a significant
overall consensus in ratings, although \citet{newcomb1956pia} reports
that three specific subjects were generally disliked by everyone,
including each other, so perhaps failing to detect this is a result of
lack of power and presence of the local nonconformity term. Notably, the
estimated correlation between the local and global nonconformity
parameters (within each week's fit) is strongly negative and
consistent ($-0.87$--$-0.77$). This suggests that these terms explain
similar behavior, though, significance of one and not the other
indicates that conformity is primarily to those more liked.

\subsubsection{Dynamic Analysis}

We now turn to modeling the evolution of the rankings over time. For
our dynamic analysis, we use a simple Markov formulation similar to
those of \citet{krackhardt2007hvs} and \citet{hanneke2010dtm} for the evolution of
rankings over time:
\[
  \Pteg(\Yyvat{}|\Yyvat{-1})=\frac{\exp\en\{\}{\innerprod{\paramv}{\genstats(\yvat{};\yvat{-1})}}}{\ceg(\paramv,\yvat{-1})},
\]
 having the normalizing constant
\[\ceg(\paramv,\yvat{-1})=\sum_\ypnetsY\exp\en\{\}{\innerprod{\paramv}{\genstats(\yv';\yvat{-1})}}.\]
For each of the 14 \emph{transitions} between successive networks, in addition to the three terms used in the cross-sectional analysis, we model inconsistency over time via $\genstat{I}(\Yvat{};\Yvat{-1})$ \eqref{eq:inconsistency}.
This term effectively absorbs the structure
in the network at $t$ that is due to inertia and to social forces
operating in the time periods prior to $t-1$, and thus the other terms
model the social forces affecting only the \emph{changes} in the
rankings over time.

Because we seek to examine the strengths of the factors over time, we
use time-varying parameters (although \citet{krivitsky2010smd} show
that the approach of \citet{hunter2006ice} can be applied to series of
networks or transitions as well). Week 9 rankings were not
reported. Because of this, for Week 10, we fit the parameters for
transition from Week 8.

\begin{figure}
  \centerline{\includegraphics[width=1\textwidth]{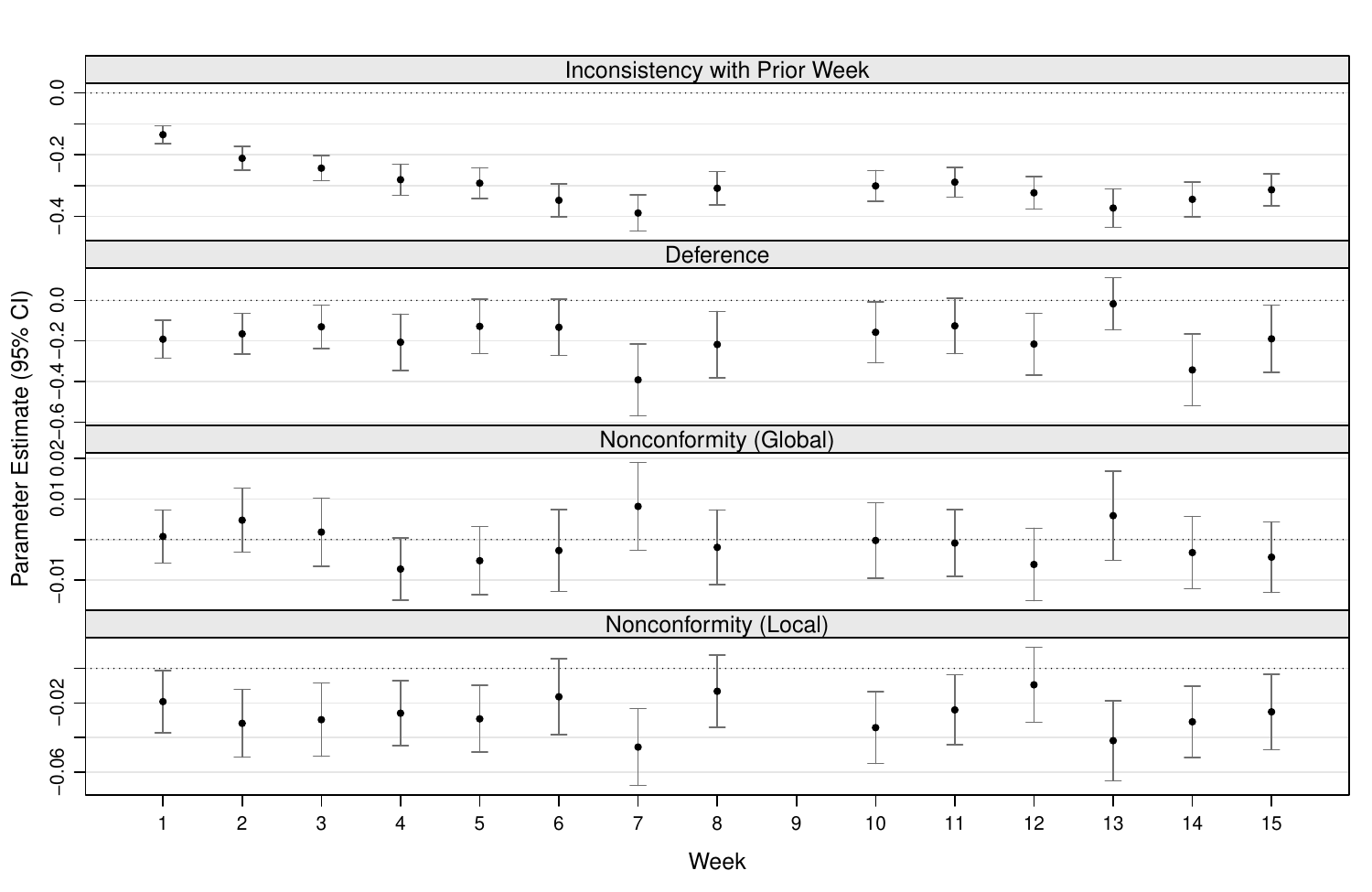}}
  \caption{\label{fig:long-coefs}Estimated coefficients for the
    longitudinal model fit to each week's rankings in the \citeauthor{newcomb1961ap}'s
    Fraternity. Error bars are at 95\% confidence.}
\end{figure}
The maximum likelihood estimates for each transition are reported in
Table~\ref{tab:long-coefs} and
visualized in Figure~\ref{fig:long-coefs}.  The estimates for the
transition from Week 8 to Week 10 do not appear to be qualitatively
different from those from nearby transitions. In particular,
inconsistency does not appear to be higher over this particular
two-week period.
\begin{table}
\begin{center}
\caption{\label{tab:long-coefs}Results for dynamic analysis of the \citeauthor{newcomb1961ap}'s data }
{ \small
\begin{tabular}{crrrr}
\hline
 & \multicolumn{4}{c}{Estimates (Std. Errors)} \\
\cline{2-5}
\centercol{Week} & \centercol{Inconsistency}  & & \multicolumn{2}{c}{Nonconformity}   \\
\cline{4-5}
\centercol{Transition} & \centercol{with Prior Week} & \centercol{Deference} & \centercol{Global} & \centercol{Local}  \\
\hline
$\phantom{1}0\to 1\phantom{0}$ & \coef{-0.135}{0.015}{\star\star\star} & \coef{-0.192}{0.048}{\star\star\star} & \coef{0.001}{0.003}{\phantom{\star}\phantom{\star}\phantom{\star}} & \coef{-0.019}{0.009}{\star\phantom{\star}\phantom{\star}} \\
$\phantom{1}1\to 2\phantom{0}$ & \coef{-0.212}{0.020}{\star\star\star} & \coef{-0.165}{0.051}{\star\star\phantom{\star}} & \coef{0.005}{0.004}{\phantom{\star}\phantom{\star}\phantom{\star}} & \coef{-0.032}{0.010}{\star\star\phantom{\star}} \\
$\phantom{1}2\to 3\phantom{0}$ & \coef{-0.244}{0.021}{\star\star\star} & \coef{-0.130}{0.055}{\star\phantom{\star}\phantom{\star}} & \coef{0.002}{0.004}{\phantom{\star}\phantom{\star}\phantom{\star}} & \coef{-0.030}{0.011}{\star\star\phantom{\star}} \\
$\phantom{1}3\to 4\phantom{0}$ & \coef{-0.281}{0.026}{\star\star\star} & \coef{-0.206}{0.071}{\star\star\phantom{\star}} & \coef{-0.007}{0.004}{\phantom{\star}\phantom{\star}\phantom{\star}} & \coef{-0.026}{0.010}{\star\star\phantom{\star}} \\
$\phantom{1}4\to 5\phantom{0}$ & \coef{-0.292}{0.026}{\star\star\star} & \coef{-0.128}{0.068}{\phantom{\star}\phantom{\star}\phantom{\star}} & \coef{-0.005}{0.004}{\phantom{\star}\phantom{\star}\phantom{\star}} & \coef{-0.029}{0.010}{\star\star\phantom{\star}} \\
$\phantom{1}5\to 6\phantom{0}$ & \coef{-0.348}{0.027}{\star\star\star} & \coef{-0.133}{0.071}{\phantom{\star}\phantom{\star}\phantom{\star}} & \coef{-0.003}{0.005}{\phantom{\star}\phantom{\star}\phantom{\star}} & \coef{-0.016}{0.011}{\phantom{\star}\phantom{\star}\phantom{\star}} \\
$\phantom{1}6\to 7\phantom{0}$ & \coef{-0.389}{0.030}{\star\star\star} & \coef{-0.392}{0.091}{\star\star\star} & \coef{0.008}{0.006}{\phantom{\star}\phantom{\star}\phantom{\star}} & \coef{-0.046}{0.011}{\star\star\star} \\
$\phantom{1}7\to 8\phantom{0}$ & \coef{-0.309}{0.028}{\star\star\star} & \coef{-0.218}{0.083}{\star\star\phantom{\star}} & \coef{-0.002}{0.005}{\phantom{\star}\phantom{\star}\phantom{\star}} & \coef{-0.013}{0.011}{\phantom{\star}\phantom{\star}\phantom{\star}} \\
$\phantom{1}8\to 10$ & \coef{-0.301}{0.026}{\star\star\star} & \coef{-0.157}{0.077}{\star\phantom{\star}\phantom{\star}} & \coef{-0.000}{0.005}{\phantom{\star}\phantom{\star}\phantom{\star}} & \coef{-0.034}{0.011}{\star\star\phantom{\star}} \\
$10\to 11$ & \coef{-0.289}{0.025}{\star\star\star} & \coef{-0.126}{0.070}{\phantom{\star}\phantom{\star}\phantom{\star}} & \coef{-0.001}{0.004}{\phantom{\star}\phantom{\star}\phantom{\star}} & \coef{-0.024}{0.010}{\star\phantom{\star}\phantom{\star}} \\
$11\to 12$ & \coef{-0.324}{0.027}{\star\star\star} & \coef{-0.216}{0.078}{\star\star\phantom{\star}} & \coef{-0.006}{0.005}{\phantom{\star}\phantom{\star}\phantom{\star}} & \coef{-0.009}{0.011}{\phantom{\star}\phantom{\star}\phantom{\star}} \\
$12\to 13$ & \coef{-0.373}{0.032}{\star\star\star} & \coef{-0.017}{0.066}{\phantom{\star}\phantom{\star}\phantom{\star}} & \coef{0.006}{0.006}{\phantom{\star}\phantom{\star}\phantom{\star}} & \coef{-0.042}{0.012}{\star\star\star} \\
$13\to 14$ & \coef{-0.345}{0.029}{\star\star\star} & \coef{-0.343}{0.090}{\star\star\star} & \coef{-0.003}{0.005}{\phantom{\star}\phantom{\star}\phantom{\star}} & \coef{-0.031}{0.010}{\star\star\phantom{\star}} \\
$14\to 15$ & \coef{-0.314}{0.027}{\star\star\star} & \coef{-0.190}{0.085}{\star\phantom{\star}\phantom{\star}} & \coef{-0.004}{0.004}{\phantom{\star}\phantom{\star}\phantom{\star}} & \coef{-0.025}{0.011}{\star\phantom{\star}\phantom{\star}} \\
\hline
\multicolumn{3}{l}{Significance levels: $0.05 \ge ^{\star} > 0.01 \ge ^{\star\star} > 0.001 \ge ^{\star\star\star}$}
\end{tabular}
}
\end{center}
\end{table}

The clear downward trend ($\hat{\tau}=-0.54$, $P\text{-value}=0.007$) in
inconsistency over successive weeks suggests that the rankings are
initially in flux as the acquaintance process takes place, solidifying
over time. As before, global nonconformity is not a significant factor. The
parameter estimates for the other two factors still appear to be, on
the whole, significant, but are uniformly smaller in magnitude
compared to those of the corresponding weeks in the cross-sectional
analysis and are less precisely estimated (as represented by uniformly
greater standard errors). This is because rather than embodying the
structure of the whole network, they embody only the structure of
\emph{changes} in the network over the week, and information to infer
their strength is drawn only from those changes. This means that some
\enquote{instant} social effects such as friendship reciprocation have
been absorbed into the Week 0 observation, which is not modeled in the
dynamic analysis.

In contrast to the cross-sectional analysis, neither deference nor
local nonconformity appear to have a significant monotone trend over
time (both correlations have $P\text{-value}\ge 0.5$). That is, while
they are time-varying when viewed cross-sectionally, the effects of
these social forces over and above inertia are fairly consistent over
time, at least after the initial time point. This suggests that this
modeling approach may be successfully isolating social forces as they
affect actors' behavior over time from the effects of preexisting
configurations.  

\subsection{\label{sec:bk}Informant Accuracy}
In the late 1970s, \citet{bernard1984pia} conducted a series of
studies to assess the accuracy of retrospective sociometric surveys of
several types. In each study, respondents in a social network---deaf
teletype users; amateur radio operators; office workers at a firm; students in a fraternity; and
faculty, graduate students, and staff in an academic program---had their social interactions observed or
recorded and were asked, in retrospect, to indicate others in their
network with whom they interacted, allowing recalled and observed
network structures to be compared. In the latter study---conducted
in a graduate program in technology education at West Virginia
University---the 34 subjects had the frequency of their interactions
recorded by a team of observers over the course of a week, and then
each subject was asked to provide a complete ranking of the other
subjects on \enquote{most to least communication that
  week} \citep{bernard1977ias}. This produces a complete ranking, suitable for analysis using our methods.

\subsubsection{Modeling inconsistency}

In this application, we use models with
sufficient statistics of the form \eqref{eq:inconsistency} and
\eqref{eq:wt-inconsistency} to assess factors that appear to affect
accuracy of rankings. Let $\yv\in \netsY$ be the reported rankings; and
let $\yv^{\text{ct.}}\in \naturals_0^\dysY$ be a weighted symmetric graph of the
observed frequencies of interaction, so that $\y^{\text{ct.}}\sij$ is
the number of times $i$ and $j$ were observed interacting. Like
$\yvat{-1}$ in the previous example, it is exogenous in our framework.

For notational convenience, we reexpress \eqref{eq:rankergm} and
\eqref{eq:wt-inconsistency} as
\begin{subequations}\label{eq:wt-inconsistency-ergm}
\begin{multline}
  \Ptegx(\Yy|\Yv^{\text{ct.}}=\yv^{\text{ct.}})=\\
\frac{\exp\EN[]{\sum_{(i,j,k)\in\distuples{3}} \en\{\}{\ypref{i}{j}{k}\en(){1-\egopref{\y^{\text{ct.}}}{i}{j}{k}} + \en(){1-\ypref{i}{j}{k}} \egopref{\y^{\text{ct.}}}{i}{j}{k} }\w\sijk(\paramv;\xv)}}{\cegx(\paramv)}, 
\end{multline}
where
\begin{equation}
\w\sijk(\paramv;\xv)=\innerprod{\paramv}{\xv\sijk},\label{eq:wt-inconsistency-ergm-w}
\end{equation}
\end{subequations}
for covariate \newtxt{$\nactors\times\nactors\times\nactors\times\npar$-}array $\xv \in \reals^{\distuples{3}\times
  \fromthru{1}{\npar}}$, so that $\xv\sijk$ is the
$\npar$-vector of covariates associated with $i$'s comparison of
$j$ and $k$, and, as before,
$\xv\sijk\equiv\xv_{i,k,j}$.  This allows us to model
inconsistency in a form semblant of logistic regression. Unlike
logistic regression for accuracy of pairwise comparisons, this model
takes into account the dependence between the comparisons that is
induced by the structure of the sample space. (I.e., that
$\ypref{i}{j}{k}\land \ypref{i}{k}{\l} \implies \ypref{i}{j}{\l}$.)

Because there are \enquote{ties} among the observed interaction
frequencies (\newtxt{i.e., where $\y^{\text{ct.}}\sij=\y^{\text{ct.}}\sik$, so $i$ interacted with $j$ and $k$ equally often}),
while the reported rankings are forced to be complete (no ties were
allowed), there is no configuration of rankings $\ynetsY$, such that
statistic \eqref{eq:inconsistency} is $0$---that the reported
rankings are completely consistent with those observed. This reveals
an interesting property of the proposed class of models: because the
comparisons that are tied in the observed frequencies simply add a
constant to their sufficient statistic, their effect on the
likelihood is canceled by the normalizing constant. That is,
inconsistent comparisons only affect the model and the estimation if
it is possible for them to be consistent is in the sample space.

For convenience, further let
$\y^{\text{obs.}}\sij$ be the \emph{observed} rank of $j$ among those
with whom $i$ had interacted, with $33$ being code for the
highest frequency, $1$ being code for the lowest frequency,
and ranks for tied alters (i.e., $\y^{\text{ct.}}\sij=\y^{\text{ct.}}\sik$, for $k\ne j$) being
computed by averaging the ranks \newtxt{that} these alters
share.

\subsubsection{Effect of Frequency of Interaction}

The first question we address is whether the magnitude of the
difference in the frequency of interaction affects the accuracy. That
is, if $i$'s frequency of interaction with $j$ differs from $i$'s
frequency of interactions with $k$ by more than $i$'s frequency of
interaction with $j$ differs from $i$'s frequency of interaction with
$\l$, is $i$ more likely to rank $j$ and $k$ accurately than $j$ and
$\l$?

To answer this, we begin by fitting a simple model with two
covariates: $\x_{i,j,k,1}=1$ in the form
\eqref{eq:wt-inconsistency-ergm}, equivalent to plain inconsistency
\eqref{eq:inconsistency}; and
$\x_{i,j,k,2}=\abs{\y^{\text{ct.}}\sij-\y^{\text{ct.}}\sik}$, the absolute difference
between the interaction frequency of $i$ with $j$ and $k$. We report
the results in Table~\ref{tab:bk-freqdiff-coefs}. Greater difference
in interaction frequency of two alters does appear to lead to greater
accuracy (i.e., lower inaccuracy) in reporting their relative ranks.
\begin{table}
\begin{center}
\caption{\label{tab:bk-freqdiff-coefs} Effect of frequency of interaction on reporting inaccuracy  }
{ \small
\begin{tabular}{lrr}
\hline
 & \multicolumn{2}{c}{Estimates (Std. Errors)} \\
\cline{2-3}
Term & \centercol{Frequency} & \centercol{Rank} \\
\hline
Intercept & \coef{-0.066}{0.020}{\star\star\phantom{\star}} & \coef{-0.159}{0.023}{\star\star\star}\\
Frequency difference & \coef{-0.018}{0.003}{\star\star\star}&  \\
Frequency rank difference & & \coef{-0.003}{0.003}{\phantom{\star}\phantom{\star}\phantom{\star}} \\
\hline
\multicolumn{3}{l}{Significance levels: $0.05 \ge ^{\star} > 0.01 \ge ^{\star\star} > 0.001 \ge ^{\star\star\star}$}
\end{tabular}
}
\end{center}
\end{table}

We also fit a similar model where we replace frequency difference with
frequency rank difference:
$\x_{i,j,k,2}=\abs{\y^{\text{obs.}}\sij-\y^{\text{obs.}}\sik}$. We find that the
effect of rank difference is, as expected, negative, but not
statistically significantly so. This lack of significance may be
counterintuitive, but it is, in fact, a consequence of the constraints
imposed by the sample space. Intuitively, given that $i$ ranks $j$ and
$k$ adjacently, an inaccurate reporting of the pairwise comparison of
these alters (e.g., $\egopref{\y^{\text{ct.}}}{i}{j}{k}$ but $\ypref{i}{k}{j}$)
does not entail misreporting any other pairwise comparisons,
including those involving $j$ or $k$; but if $j$ and $k$ are some
$d>1$ ranks apart in $\yv^{\text{ct.}}$---they have $d-1$ other alters between
them---then inaccurately reporting the pairwise comparison of
$j$ and $k$ entails inaccurately reporting the pairwise comparisons
$\egopref{\y^{\text{ct.}}}{i}{j}{\l}$ and/or $\egopref{\y^{\text{ct.}}}{i}{k}{\l}$ for every
$\l$ ranked between $j$ and $k$. It can be shown easily that any
configuration $\yv$ in which $\ypref{i}{k}{j}\ne\egopref{\y^{\text{ct.}}}{i}{j}{k}$
must also misreport at least $d-1$ such comparisons. Thus, even a
model with no rank difference effect and only a baseline inconsistency
effect would already heavily penalize inaccurate reporting of
comparisons between distantly-ranked alters.

\subsubsection{Effect of Salience}

The second question that we address is whether the accuracy of
reported ranking is affected by the positions of those being
ranked. Can an ego $i$ better discern the ranking of those with whom
\xe{} interacts the most? Is \xe{} more accurate at the extremes?

To answer this, we fit a model for inconsistency that is a quadratic
polynomial in rank values. More concretely, in the form
\eqref{eq:wt-inconsistency-ergm}, $\w\sijk(\paramv;\xv)$ has the
covariate vector
\begin{equation}\xv\sijk=\left[1,~\y^{\text{obs.}}\sij+\y^{\text{obs.}}\sik,~(\y^{\text{obs.}}\sij)^2+(\y^{\text{obs.}}\sik)^2,~\y^{\text{obs.}}\sij\y^{\text{obs.}}\sik\right],\label{eq:bk-sal-cov}\end{equation}
inducing a model in which the inconsistency of reported with observed
is modeled as a second-degree polynomial function of the observed
ranks of the alters being compared; and this function is symmetric for
these two alters. The statistics in \eqref{eq:bk-sal-cov} represent
baseline inconsistency, linear effect of ranks of the alters being
compared, their quadratic effect, and their interaction effect,
respectively.

In fitting this model, we found that it suffers from collinearity,
which impedes inference, so to improve its numeric conditioning, we fit an
equivalent model, using rescaled and centered quantiles, evaluating
$\y^{\text{q.}}\sij\equiv(\y^{\text{obs.}}\sij-17)/32$ and substituting $\yv^{\text{q.}}$ for $\yv^{\text{obs.}}$ in
\eqref{eq:bk-sal-cov}.

\begin{figure}
  \centerline{\includegraphics[width=0.5\textwidth]{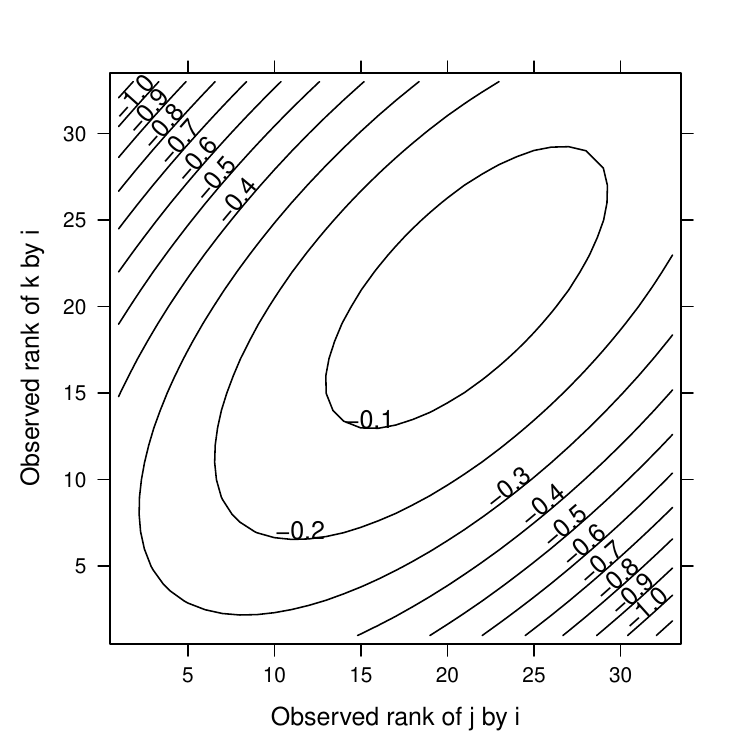}}
  \caption{\label{fig:bk-inconsistency-ranks-effect} Predicted effect of alter positions on reporting inaccuracy}
\end{figure}

We report results from this fit in
Table~\ref{tab:bk-polyrank-coefs}. All four covariates appear to be
highly significant. Somewhat surprisingly, the higher-ranked alters
appear to have slightly higher inconsistency between observed and
reported. However, the negative coefficient on the quadratic term
suggests that the middle ranks are reported with the least accuracy of
all. We show the predicted inconsistency weight
$\w\sijk(\hat{\paramv};\xv)$ as a function of their observed ranks in
Figure~\ref{fig:bk-inconsistency-ranks-effect}. From this, it appears
that, indeed, for alters whose observed rankings are close together,
the accuracy is lowest if the alters are ranked in the middle. (For
alters whose observed rankings are far apart, the accuracy is higher.)

\begin{table}
\begin{center}
\caption{\label{tab:bk-polyrank-coefs} Effects of alter positions on reporting inaccuracy  }
{ \small
\begin{tabular}{lr}
\hline
Term & Estimate (Std. Err.)\\
\hline
Intercept & \coef{-0.068}{ 0.024}{\star\star\phantom{\star}}\\
Linear & \coef{0.107}{ 0.018}{\star\star\star}\\
Quadratic & \coef{-1.300}{ 0.296}{\star\star\star}\\
Interaction & \coef{1.765}{ 0.512}{\star\star\star}\\
\hline
\multicolumn{2}{l}{Sig.: $0.05 \ge^{\star}> 0.01 \ge ^{\star\star} > 0.001 \ge ^{\star\star\star}$}
\end{tabular}
}
\end{center}
\end{table}

\section{Discussion}

\newtxt{As befits the long history of ordinal data analysis in the social sciences, many extensions and applications of the present framework are possible.  Here, we briefly discuss two such directions: the extension of the current framework to ``bipartite'' rank data; and the use of our framework in settings that imply particular kinds of ordinal constraints (e.g., partial orders, semi-orders, incompletely observed total orders, etc.).}

\subsection{Extension to ``Bipartite'' Rank Data}\label{sec:bipartite}

\newtxt{Although our focus here has been on the classic sociometric setting in which a group of individuals are asked to rank each other with respect to some dimension (e.g., liking), our modeling framework can easily accommodate other cases as well.  One such setting is the case of ``bipartite'' rank data, in which members of a given group are asked to rank a set of objects not including the group members themselves.  Examples of such data include preference rankings of political candidates, organizations, or policy positions; ordinal judgments regarding physical objects or perceptual stimuli; liking or other rankings of non-group members, etc.  Although such data is widely analyzed throughout the social sciences using traditional techniques, the contribution of our approach is the ability to model \emph{interdependence among raters} in a natural way.  For instance, the global conformity statistic of Section~\ref{sec:conformity} can be used to capture a general tendency of group members to converge on a common rating of a set of objects; likewise, the consistency effects of Section~\ref{sec:consistency} have the same meaning in the ``bipartite'' setting as in a standard sociometric setting.  The exogenous effects of Section~\ref{sec:attractiveness} can be used to capture differences in the net attractiveness of objects to members of the rating group, and dyadic covariates (Section~\ref{sec:dyadic}) can be used to measure effects related to the tendency of particular raters or subgroups of raters to give higher/lower ratings to particular objects.  On the other hand, statistics that depend upon the ratings \emph{by} the object (e.g., local conformity) are clearly meaningless in a bipartite context, and should not be employed.}

\newtxt{The bipartite case suggests certain statistics that may prove especially useful for assessing influence in structured groups.  For instance, consider the case in which our set of raters, $\egos$, interact via a known, fixed social network with adjacency matrix $\xv$.  Let $\alters$ be the set of objects to be ranked.  We may then capture the tendency for those adjacent in $\xv$ to rate objects in $\alters$ (dis)similarly via the \emph{dyadic nonconformity statistic},
\begin{equation*}
\genstat{DNC}(\yv) = \sum_{(i,\l)\in \egodistuples{2}} \x\sil \sum_{(j,k)\in \altdistuples{2}}\ypref{\l}{j}{k} \left(1-\ypref{i}{j}{k}\right). \label{eq:influence}
\end{equation*}
To the extent that adjacent actors influence each other to form similar views of the object set, $\genstat{DNC}$ will be suppressed; the associated parameter is hence analogous (up to a sign change) to the autocorrelation parameter in a standard linear network autocorrelation model (LNAM) \citep{cliff.ord:bk:1973,doreian:ch:1990}.  As with the LNAM, the adjacency structure (analogous to the weight matrix) need not be dichotomous, and can contain continuous measures of exposure, proximity, similarity, group co-membership, or the like.  Also like the LNAM, the adjacency/weight matrix is taken to be exogenous and fixed; models in which object ratings and interpersonal relationships co-evolve would be an promising direction for future research in this area.}

\newtxt{Simulation and estimation for bipartite rank data requires modifying the support of the associated model to include only the observable ranks, and eliminating impermissible rating triads from the proposals in Algorithm~\ref{alg:rank-MH} in the Appendix.  These are straightforward changes to the base implementation, and are not discussed in detail here.}  

\subsection{Considerations Relating to Types of Orderings}

It should be noted that, in our development, we focus on the case of orderings that are defined from the psychological process in question, rather than data that are ordinal simply due to limitations in measurement (e.g., count or continuous data observed only as ranks). Although the present framework may in some cases be useful for such data (e.g., to avoid having to model the count distribution), the assumptions involved, e.g., in the choice of sufficient statistics (and their interpretation), may be quite different.

This work has focused on complete orderings, but the above distinction gains further importance when considering partial orderings: the case of partial orderings being a property of underlying psychological phenomena is substantively different from the case of incomplete orderings arising from measurement itself.  A well-known example of the latter is a frequently used sociometric survey design that asks each ego to rank their top $k$ alters with respect to some criterion (liking, interaction frequency, etc.). 

\newtxt{The \citet{sampson1968npc} study is instructive in this. Recall that \citeauthor{sampson1968npc} asked each of 18 novitiates in a monastery to rank his three most liked novitiates among the other 17. That only the top three were ranked does not mean that the \emph{underlying} ordering is partial and that the remaining 14 alters are tied. Rather, their ranking relative to each other is \emph{unobserved} by design: we know that their underlying rank must be below the top three, and little else: they
 may or may not be substantively tied.}

Similarly, in 1995--1996, the Wave I of National Longitudinal Study of Adolescent Health, \newtxt{mentioned in the Introduction,} asked high school students to list, in order, up to 5 male and up to 5 female friends. \citep{harris2003nls} Again, the students unranked (either
because they fell below the respondent's threshold for friendship or
because the respondent ran out of space on the form) are not all
necessarily substantively tied (i.e., assigned the same rank) but are merely unobserved. \newtxt{In addition, because male and female students were ranked separately, there is no information about underlying ranking of each respondent's female friends relative to \xer{} male friends, creating a situation with incomparable subsets, analogous to that of Ego~$\aC$ in the toy network $\yexw$. (Notably, such data would still permit one to infer, at least for those who did not nominate their maximum of 5, that anyone (of either sex) whom they had chosen not to rank has an underlying rank lower than anyone (of either sex) whom they had chosen to rank.)}

Cases of unobserved ranking are better handled by means of a latent variable framework in which a probability distribution is placed on the complete data, and in which likelihood is assessed via the marginalization of the complete data conditional on that observed. \newtxt{(Plackett--Luce models for top-$k$ ranking data \citep{plackett1975ap} do this implicitly.)}  Such a strategy has been used in the traditional ERGM framework by \citet{handcock2010msn}, and can be employed here as well.

In contrast, orderings in which the alters may be substantively
tied or incomparable with each other---partial orderings---pose
different challenges. \newtxt{In contrast to the above, a close examination of \citeauthor{sampson1968npc}'s data reveals that some of the novitiates were, in fact, recorded assigning equal ranks to some of those they nominated; and these would be substantively tied in an underlying partial ordering. Tied ratings in data like that of \citet{johnson.et.al:jms:2003} could be interpreted either way: either that the scale of 0--10 was not sufficiently granular to encode small differences in degrees of interaction, so which of the tied alters was actually higher is unobserved; or that an ego's choice to rate two alters equally means that their degrees of interaction were substantively the same.}

 In the complete ordering case, there exists a distribution of
rank-orderings that is unambiguously uniform and can thus serve as the
baseline distribution (\emph{reference measure}) for the exponential
family \citep[see, e.g.][:115--116]{barndorffnielsen1978ief}. In
the partially ordered case, there is no natural baseline that is
unambiguously uniform. For example, an ego may rank all $\nactors-1$ alters equally (with only one such \enquote{ranking} possible) or \xe{} may assign a distinct rank to each (with $(\nactors-1)!$ possible rankings), or anything in between. Modeling of partial orderings, therefore, requires modeling not only the actors' ranking propensities but also their choices whether to rank or not to rank, and, in particular, the baseline distribution of these choices, when $\paramv=\0$. \newtxt{How to capture such behavior in a cognitively realistic way is not currently known.}

Thus, in this paper, we focused our discussion on complete
orderings. Despite our focus on complete orderings, our techniques can
(with appropriate choice of reference measure) be generalized to the
weaker case: all of the model statistics that we present in
Section~\ref{sec:terms} can be applied to partially ordered
network data without modification, because they only make references to the network of
interest through the indicator $\ypref{\cdot}{\cdot}{\cdot}$
\eqref{eq:egopref}.

\section{Conclusion}

Rank-order data are a cornerstone of sociometric measurement, but principled treatment of such data in an interpersonal context poses significant statistical challenges.  Here, we have shown how statistical exponential families may be used to generalize the now well-known ERGM framework to the rank-order case.  We have also introduced a corresponding set of sufficient statistics that are appropriate for use when only within-ego ordinal judgments are psychologically meaningful, a restriction that is important when modeling such data.  As with conventional ERGMs, a wide range of statistics may be posited to capture alternative psychological and social mechanisms; the ability to evaluate and compare competing models based on such distinct alternatives is one of the strengths of the statistical approach.

\newtxt{
Given the current wave of interest in \enquote{large} networks, computationally scalable estimation has been an increasing focus of research in the social network community.  Relative to population-scale networks, the networks considered here have been fairly small (i.e., 10s of vertices), and our focus has been on developing methods that work well on this latter scale.  This focus is motivated by the fact that sociometric data of the form treated here is typically of interest only in group or organizational settings in which all members of the network are salient and well-known to each other; since such networks are by nature fairly small (e.g., rarely larger than 20--50 persons), highly scalable techniques are less critical for rank-order models than for binary networks.  That said, computationally scalable estimation (particularly for the above-described top-$k$ design) is an interesting challenge for future research in this area.}

In our assumption that the network must be modeled through
$\ypref{\cdot}{\cdot}{\cdot}$, we have discarded any information
associated with the rank as such. \newtxt{ Where measurements are \emph{truly} ordinal, this is the appropriate choice; nevertheless, it has been noted e.g. by \citet{levine:bk:1993} that putatively ordinal data can in practice contain more information than a strict ordinal interpretation would allow.} As we show in our example
in Section~\ref{sec:bk}, rank values and rank differences can be
incorporated into the models as well, though that example only uses
them exogenously. Our framework does not, fundamentally, preclude
incorporating such effects, and, in fact, it permits separating and
testing their significance over and above the purely
comparison-based effects. This, also, is a subject for ongoing work.

Finally, we note that development of ERGMs for rank-order data opens the way to a rich family of novel statistical models for phenomena such as interdependent choice behavior in group context, social influence on preferences, etc.  Particularly because of their suitability for data collected in observational settings, rank-order ERGMs provide a useful tool for considering both new and classic problems in social psychology and the study of decision making.

\section*{Acknowledgments}
The authors wish to thank David Krackhardt for helpful
discussions. Images of characters used in Figure~1 are based (with
modifications) on clip-art authored by user Foofpurple
(\url{http://foofurple.com/mix/index.html}) and released into the
Public Domain via OpenClipart.org (\url{http://www.openclipart.org}).

\section*{Funding}
Research was supported by Portuguese Foundation for Science and
Technology Ci\^{e}ncia 2009 Program, ONR award N000140811015, and NIH
award 1R01HD068395-01. Computation and simulations were performed on a
computing cluster partially funded by a \emph{Eunice Kennedy Shriver}
National Institute of Child Health and Human Development research
infrastructure grant, R24HD042828, to the Center for Studies in
Demography and Ecology at the University of Washington.

\section*{Biography}

\paragraph{Pavel N. Krivitsky} is Lecturer in Statistics at the School
of Mathematics and Applied Statistics and the National Institute for
Applied Statistics Research Australia (NIASRA), at University of Wollongong. He
received his BS in Biometry and Statistics, Cornell University and his
PhD in Statistics at the University of Washington. His current
interest is statistical models for social network data and
processes, their formulation, implementation, and application.

\paragraph{Carter T. Butts} is Professor in the Departments of Sociology, Statistics, and Electrical Engineering and Computer Science at the University of California Irvine, where he directs the Center for Networks and Relational Analysis.  His current research interests include statistical models for network structure and dynamics, models for interaction processes, and the response of social systems to exogenous shocks.

\bibliographystyle{plainnat}
\addcontentsline{toc}{section}{References}
\bibliography{ERGMs_for_Rank-Order_Networks}

\begin{thebibliography}{60}
\providecommand{\natexlab}[1]{#1}
\providecommand{\url}[1]{\texttt{#1}}
\expandafter\ifx\csname urlstyle\endcsname\relax
  \providecommand{\doi}[1]{doi: #1}\else
  \providecommand{\doi}{doi: \begingroup \urlstyle{rm}\Url}\fi

\bibitem[Anderson et~al.(2006)Anderson, Srivastava, Beer, Spataro, and
  Chatman]{anderson.et.al:jpsp:2006}
Cameron Anderson, Sanjay Srivastava, Jennifer~S. Beer, Sandra~E. Spataro, and
  Jennifer~A. Chatman.
\newblock Knowing your place: Self-perceptions of status in face-to-face
  groups.
\newblock \emph{Journal of Personality and Social Psychology}, 91\penalty0
  (6):\penalty0 1094--1110, 2006.

\bibitem[Arabie et~al.(1978)Arabie, Boorman, and Levitt]{arabie1978cbh}
Phipps Arabie, Scott~A. Boorman, and Paul~R. Levitt.
\newblock Constructing blockmodels: {How} and why.
\newblock \emph{Journal of Mathematical Psychology}, 17\penalty0 (1):\penalty0
  21--63, 1978.
\newblock ISSN 0022-2496.
\newblock \doi{10.1016/0022-2496(78)90034-2}.

\bibitem[Asch(1951)]{asch1951egp}
Solomon~E. Asch.
\newblock \emph{Groups, Leadership and Men}, chapter Effects of Group Pressure
  Upon the Modification and Distortion of Judgments, pages 177--190.
\newblock Carnegie Press, 1951.

\bibitem[Barndorff-Nielsen(1978)]{barndorffnielsen1978ief}
Ole~E. Barndorff-Nielsen.
\newblock \emph{Information and Exponential Families in Statistical Theory}.
\newblock John Wiley \& Sons, Inc., New York, 1978.
\newblock ISBN 0471995452.

\bibitem[Berger et~al.(1972)Berger, Cohen, and Zelditch]{berger.et.al:asr:1972}
Joseph Berger, Bernard~P. Cohen, and Morris Zelditch.
\newblock Status characteristics and social interaction.
\newblock \emph{American Sociological Review}, 37\penalty0 (3):\penalty0
  241--255, 1972.

\bibitem[Berger et~al.(1977)Berger, Fisek, Norman, and
  Zelditch]{berger.et.al:bk:1977}
Joseph Berger, M.~Hamit Fisek, Robert~Z. Norman, and Morris Zelditch.
\newblock \emph{Status Characteristics and Social Interaction: An Expectation
  States Approach}.
\newblock Elsevier, New York, 1977.

\bibitem[Bernard and Killworth(1977)]{bernard1977ias}
H.~Russell Bernard and Peter~D. Killworth.
\newblock Informant accuracy in social network data {II}.
\newblock \emph{Human Communication Research}, 4\penalty0 (1):\penalty0 3--18,
  1977.
\newblock ISSN 1468-2958.
\newblock \doi{10.1111/j.1468-2958.1977.tb00591.x}.

\bibitem[Bernard et~al.(1984)Bernard, Killworth, Kronenfeld, and
  Sailer]{bernard1984pia}
H.~Russell Bernard, Peter Killworth, David Kronenfeld, and Lee Sailer.
\newblock The problem of informant accuracy: {The} validity of retrospective
  data.
\newblock \emph{Annual Review of Anthropology}, 13:\penalty0 495--517, 1984.
\newblock ISSN 0084-6570.

\bibitem[Bordieu(1968)]{bordieu:issj:1968}
Pierre Bordieu.
\newblock Outline of a sociological theory of art perception.
\newblock \emph{International Social Science Journal}, 20:\penalty0 589--612,
  1968.

\bibitem[Breiger et~al.(1975)Breiger, Boorman, and Arabie]{breiger1975acr}
Ronald~L. Breiger, Scott~A. Boorman, and Phipps Arabie.
\newblock An algorithm for clustering relational data with applications to
  social network analysis and comparison with multidimensional scaling.
\newblock \emph{Journal of Mathematical Psychology}, 12:\penalty0 323--383,
  1975.
\newblock \doi{10.1016/0022-2496(75)90028-0}.

\bibitem[Brown(1986)]{brown1986fse}
Lawrence~D. Brown.
\newblock \emph{Fundamentals of Statistical Exponential Families with
  Applications in Statistical Decision Theory}, volume~9 of \emph{Lecture
  Notes---Monograph Series}.
\newblock Institute of Mathematical Statistics, Hayward, California, 1986.
\newblock ISBN 0-940600-10-2.
\newblock URL \url{http://projecteuclid.org/euclid.lnms/1215466757}.

\bibitem[Butts(2007{\natexlab{a}})]{butts2007pmr}
Carter~T. Butts.
\newblock Permutation models for relational data.
\newblock \emph{Sociological Methodology}, 37\penalty0 (1):\penalty0 257--281,
  2007{\natexlab{a}}.
\newblock \doi{10.1111/j.1467-9531.2007.00183.x}.

\bibitem[Butts(2007{\natexlab{b}})]{butts:sm:2007b}
Carter~T. Butts.
\newblock Models for generalized location systems.
\newblock \emph{Sociological Methodology}, 37\penalty0 (1):\penalty0 283--348,
  2007{\natexlab{b}}.
\newblock \doi{10.1111/j.1467-9531.2006.00187.x}.

\bibitem[Butts(2011)]{butts2011bgb}
Carter~T. Butts.
\newblock Bernoulli graph bounds for general random graphs.
\newblock \emph{Sociological Methodology}, 41\penalty0 (1):\penalty0 299--345,
  2011.
\newblock \doi{10.1111/j.1467-9531.2011.01246.x}.

\bibitem[Caimo and Friel(2011)]{CaFr11b}
Alberto Caimo and Nial Friel.
\newblock Bayesian inference for exponential random graph models.
\newblock \emph{Social Networks}, 33\penalty0 (1):\penalty0 41--55, 2011.
\newblock \doi{10.1016/j.socnet.2010.09.004}.

\bibitem[Cliff and Ord(1973)]{cliff.ord:bk:1973}
Andrew~D. Cliff and J.~Keith Ord.
\newblock \emph{Spatial Autocorrelation}.
\newblock Pion, London, 1973.

\bibitem[Dekker et~al.(2007)Dekker, Krackhardt, and
  Snijders]{dekker.et.al:p:2007}
David Dekker, David Krackhardt, and Tom A.~B. Snijders.
\newblock Sensitivity of {MRQAP} tests to collinearity and autocorrelation
  conditions.
\newblock \emph{Psychometrika}, 72\penalty0 (4):\penalty0 563--581, 2007.

\bibitem[Doreian(1990)]{doreian:ch:1990}
P.~Doreian.
\newblock Network autocorrelation models: Problems and prospects.
\newblock In I.~D.~A. Griffith, editor, \emph{Spatial Statistics: Past,
  Present, and Future}, pages 369--389. Institute of Mathematical Geography,
  Ann Arbor, 1990.

\bibitem[Doreian et~al.(1996)Doreian, Kapuscinski, Krackhardt, and
  Szczypula]{doreian1996bhb}
Patrick Doreian, Roman Kapuscinski, David Krackhardt, and Janusz Szczypula.
\newblock A brief history of balance through time.
\newblock \emph{Journal of Mathematical Sociology}, 21\penalty0
  (1--2):\penalty0 113--131, 1996.

\bibitem[Dunbar(1997)]{dunbar:bk:1997}
Robin Dunbar.
\newblock \emph{Grooming, Gossip, and the Evolution of Language}.
\newblock Harvard University Press, Cambridge, MA, 1997.

\bibitem[Enelow and Hinich(1984)]{enelow.hinich:bk:1984}
James~M. Enelow and Melvin~J. Hinich.
\newblock \emph{The Spatial Theory of Voting: An Introduction}.
\newblock Cambridge, Cambridge, 1984.

\bibitem[Geyer and Thompson(1992)]{geyer1992cmc}
Charles~J. Geyer and Elizabeth~A. Thompson.
\newblock Constrained {Monte} {Carlo} maximum likelihood for dependent data
  (with discussion).
\newblock \emph{Journal of the Royal Statistical Society. Series B},
  54\penalty0 (3):\penalty0 657--699, 1992.
\newblock ISSN 0035-9246.

\bibitem[Goodreau et~al.(2008{\natexlab{a}})Goodreau, Handcock, Hunter, Butts,
  and Morris]{goodreau2008st}
Steven~M. Goodreau, Mark~S. Handcock, David~R. Hunter, Carter~T. Butts, and
  Martina Morris.
\newblock A \pkg{statnet} tutorial.
\newblock \emph{Journal of Statistical Software}, 24\penalty0 (9):\penalty0
  1--26, May 2008{\natexlab{a}}.
\newblock ISSN 1548-7660.
\newblock URL \url{http://www.jstatsoft.org/v24/i09}.

\bibitem[Goodreau et~al.(2008{\natexlab{b}})Goodreau, Kitts, and
  Morris]{goodreau2008bff}
Steven~M. Goodreau, James Kitts, and Martina Morris.
\newblock Birds of a feather, or friend of a friend? {Using} exponential random
  graph models to investigate adolescent social networks.
\newblock \emph{Demography}, 45\penalty0 (1):\penalty0 103--125, February
  2008{\natexlab{b}}.
\newblock ISSN 0070-3370.

\bibitem[Gormley and Murphy(2008)]{gormley2008evb}
Isobel~Claire Gormley and Thomas~Brendan Murphy.
\newblock Exploring voting blocs within the {Irish} electorate: {A} mixture
  modeling approach.
\newblock \emph{Journal of the American Statistical Association}, 103\penalty0
  (483):\penalty0 1014--1027, 2008.
\newblock ISSN 0162-1459.
\newblock \doi{10.1198/016214507000001049}.

\bibitem[Handcock and Gile(2010)]{handcock2010msn}
Mark~S. Handcock and Krista~J. Gile.
\newblock Modeling social networks from sampled data.
\newblock \emph{Annals of Applied Statistics}, 4\penalty0 (1):\penalty0 5--25,
  2010.
\newblock ISSN 1932-6157.
\newblock \doi{10.1214/08-AOAS221}.

\bibitem[Handcock et~al.(2014)Handcock, Hunter, Butts, Goodreau, Krivitsky, and
  Morris]{handcock2010epf}
Mark~S. Handcock, David~R. Hunter, Carter~T. Butts, Steven~M. Goodreau,
  Pavel~N. Krivitsky, and Martina Morris.
\newblock \emph{\pkg{ergm}: Fit, Simulate and Diagnose Exponential-Family
  Models for Networks}.
\newblock The Statnet Project (\url{http://www.statnet.org}), 2014.
\newblock URL \url{CRAN.R-project.org/package=ergm}.
\newblock R package version 3.2.4.

\bibitem[Hanneke et~al.(2010)Hanneke, Fu, and Xing]{hanneke2010dtm}
Steve Hanneke, Wenjie Fu, and Eric~P. Xing.
\newblock Discrete temporal models of social networks.
\newblock \emph{Electronic Journal of Statistics}, 4:\penalty0 585--605, 2010.
\newblock ISSN 1935-7524.
\newblock \doi{10.1214/09-EJS548}.

\bibitem[Harris et~al.(2003)Harris, Florey, Tabor, Bearman, Jones, and
  Udry]{harris2003nls}
Kathleen~M. Harris, F.~Florey, Joyce Tabor, Peter~S. Bearman, J.~Jones, and
  J.~Richard Udry.
\newblock The national longitudinal study of adolescent health: Research
  design.
\newblock Technical report, University of North Carolina, 2003.
\newblock URL \url{http://www.cpc.unc.edu/projects/addhealth/design/}.

\bibitem[Holland and Leinhardt(1981)]{holland1981efp}
Paul~W. Holland and Samuel Leinhardt.
\newblock An exponential family of probability distributions for directed
  graphs.
\newblock \emph{Journal of the American Statistical Association}, 76\penalty0
  (373):\penalty0 33--65, 1981.
\newblock ISSN 0162-1459.

\bibitem[Hubert(1987)]{hubert:bk:1987}
Lawrence~J. Hubert.
\newblock \emph{Assignment Methods in Combinatorial Data Analysis}.
\newblock Marcel Dekker, New York, 1987.

\bibitem[Hummel et~al.(2012)Hummel, Hunter, and Handcock]{HuHu12i}
Ruth~M. Hummel, David~R. Hunter, and Mark~S. Handcock.
\newblock Improving simulation-based algorithms for fitting {ERGMs}.
\newblock \emph{Journal of Computational and Graphical Statistics}, 21\penalty0
  (4):\penalty0 920--939, 2012.
\newblock \doi{10.1080/10618600.2012.679224}.

\bibitem[Hunter and Handcock(2006)]{hunter2006ice}
David~R. Hunter and Mark~S. Handcock.
\newblock Inference in curved exponential family models for networks.
\newblock \emph{Journal of Computational and Graphical Statistics}, 15\penalty0
  (3):\penalty0 565--583, 2006.
\newblock ISSN 1061-8600.

\bibitem[Hunter et~al.(2008{\natexlab{a}})Hunter, Goodreau, and
  Handcock]{hunter2008gfs}
David~R. Hunter, Steven~M. Goodreau, and Mark~S. Handcock.
\newblock Goodness of fit for social network models.
\newblock \emph{Journal of the American Statistical Association}, 103\penalty0
  (481):\penalty0 248--258, March 2008{\natexlab{a}}.
\newblock ISSN 0162-1459.
\newblock \doi{10.1198/016214507000000446}.

\bibitem[Hunter et~al.(2008{\natexlab{b}})Hunter, Handcock, Butts, Goodreau,
  and Morris]{hunter2008epf}
David~R. Hunter, Mark~S. Handcock, Carter~T. Butts, Steven~M. Goodreau, and
  Martina Morris.
\newblock \pkg{ergm}: A package to fit, simulate and diagnose
  exponential-family models for networks.
\newblock \emph{Journal of Statistical Software}, 24\penalty0 (3):\penalty0
  1--29, May 2008{\natexlab{b}}.
\newblock ISSN 1548-7660.
\newblock URL \url{http://www.jstatsoft.org/v24/i03}.

\bibitem[Johnson et~al.(2003)Johnson, Boster, and
  Palinkas]{johnson.et.al:jms:2003}
Jeffrey~C. Johnson, James~S. Boster, and Lawrence~A. Palinkas.
\newblock Social roles and the evolution of networks.
\newblock \emph{Journal of Mathematical Sociology}, 27:\penalty0 89--121, 2003.

\bibitem[Krackhardt(1987)]{krackhardt:sn:1987}
David Krackhardt.
\newblock {QAP} partialling as a test of spuriousness.
\newblock \emph{Social Networks}, 9\penalty0 (2):\penalty0 171--186, 1987.

\bibitem[Krackhardt and Handcock(2007)]{krackhardt2007hvs}
David Krackhardt and Mark~S. Handcock.
\newblock {Heider} versus {Simmel}: {Emergent} features in dynamic structures.
\newblock In Edoardo Airoldi, David~M. Blei, Stephen~E. Fienberg, Anna
  Goldenberg, Eric~P. Xing, and Alice~X. Zheng, editors, \emph{Statistical
  Network Analysis: Models, Issues, and New Directions: ICML 2006 Workshop on
  Statistical Network Analysis, Pittsburgh, PA, USA, June 29, 2006, Revised
  Selected Papers}, volume 4503 of \emph{Lecture Notes in Computer Science},
  pages 14--27. Springer, 2007.
\newblock ISBN 978-3-540-73132-0.
\newblock \doi{10.1007/978-3-540-73133-7_2}.

\bibitem[Krivitsky(2012)]{krivitsky2012erg}
Pavel~N. Krivitsky.
\newblock Exponential-family random graph models for valued networks.
\newblock \emph{Electronic Journal of Statistics}, 6:\penalty0 1100--1128,
  2012.
\newblock \doi{10.1214/12-EJS696}.

\bibitem[Krivitsky and Handcock(2014)]{krivitsky2010smd}
Pavel~N. Krivitsky and Mark~S. Handcock.
\newblock A separable model for dynamic networks.
\newblock \emph{Journal of the Royal Statistical Society, Series B},
  76\penalty0 (1):\penalty0 29--46, 2014.
\newblock \doi{10.1111/rssb.12014}.

\bibitem[Levine(1993)]{levine:bk:1993}
Joel~H. Levine.
\newblock \emph{Exceptions Are the Rule: Inquiries on Method in the Social
  Sciences}.
\newblock Westview Press, 1993.

\bibitem[Morse et~al.(1974)Morse, Reis, Gruzen, and Wolff]{morse.et.al:jp:1974}
Stanley~J. Morse, Harry~T. Reis, Joan Gruzen, and Ellen Wolff.
\newblock The eye of the beholder: {Determinants} of physical attractiveness
  judgments in the {U.S.} and {South} {Africa}.
\newblock \emph{Journal of Personality}, 42:\penalty0 528--542, 1974.

\bibitem[Nakao and Romney(1993)]{nakao1993las}
Keiko Nakao and A.~Kimball Romney.
\newblock Longitudinal approach to subgroup formation: Re-analysis of
  {Newcomb's} fraternity data.
\newblock \emph{Social Networks}, 15\penalty0 (2):\penalty0 109--131, June
  1993.
\newblock ISSN 0378-8733.
\newblock \doi{10.1016/0378-8733(93)90001-2}.

\bibitem[Newcomb(1956)]{newcomb1956pia}
Theodore~M. Newcomb.
\newblock The prediction of interpersonal attraction.
\newblock \emph{American Psychologist}, 11\penalty0 (11):\penalty0 575--586,
  1956.
\newblock ISSN 0003-066X.
\newblock \doi{10.1037/h0046141}.

\bibitem[Newcomb(1961)]{newcomb1961ap}
Theodore~M. Newcomb.
\newblock \emph{The Acquaintance Process}.
\newblock Holt, Rinehart, Winston., New York, 1961.

\bibitem[Pattison(1982)]{pattison1982asm}
Philippa~E. Pattison.
\newblock The analysis of semigroups of multirelational systems.
\newblock \emph{Journal of Mathematical Psychology}, 25\penalty0 (2):\penalty0
  87--118, 1982.
\newblock ISSN 0022-2496.
\newblock \doi{10.1016/0022-2496(82)90008-6}.

\bibitem[Plackett(1975)]{plackett1975ap}
Robin~L. Plackett.
\newblock The analysis of permutations.
\newblock \emph{Journal of the Royal Statistical Society. Series C (Applied
  Statistics)}, 24\penalty0 (2):\penalty0 193--202, 1975.
\newblock ISSN 0035-9254.

\bibitem[{R Core Team}(2014)]{RC14r}
{R Core Team}.
\newblock \emph{\proglang{R}: A Language and Environment for Statistical
  Computing}.
\newblock R Foundation for Statistical Computing, Vienna, Austria, 2014.
\newblock URL \url{http://www.R-project.org/}.

\bibitem[Rinaldo et~al.(2009)Rinaldo, Fienberg, and Zhou]{rinaldo2009gde}
Alessandro Rinaldo, Stephen~E. Fienberg, and Yi~Zhou.
\newblock On the geometry of discrete exponential families with application to
  exponential random graph models.
\newblock \emph{Electronic Journal of Statistics}, 3:\penalty0 446--484, 2009.
\newblock ISSN 1935-7524.
\newblock \doi{10.1214/08-EJS350}.

\bibitem[Robins et~al.(1999)Robins, Pattison, and Wasserman]{robins1999lml}
Garry Robins, Philippa Pattison, and Stanley~S. Wasserman.
\newblock Logit models and logistic regressions for social networks: {III.}
  {Valued} relations.
\newblock \emph{Psychometrika}, 64\penalty0 (3):\penalty0 371--394, 1999.
\newblock ISSN 0033-3123.

\bibitem[Sampson(1968)]{sampson1968npc}
Samuel~F. Sampson.
\newblock \emph{A Novitiate in a Period of Change: {An} Experimental and Case
  Study of Social Relationships}.
\newblock {Ph.D.}\ thesis (university micofilm, no 69-5775), Department of
  Sociology, Cornell University, Ithaca, New York, 1968.

\bibitem[Schweinberger(2011)]{schweinberger2011isd}
Michael Schweinberger.
\newblock Instability, sensitivity, and degeneracy of discrete exponential
  families.
\newblock \emph{Journal of the American Statistical Association}, 106\penalty0
  (496):\penalty0 1361--1370, 2011.
\newblock \doi{10.1198/jasa.2011.tm10747}.

\bibitem[Snijders et~al.(2006)Snijders, Pattison, Robins, and
  Handcock]{snijders2006nse}
Tom A.~B. Snijders, Philippa~E. Pattison, Garry~L. Robins, and Mark~S.
  Handcock.
\newblock New specifications for exponential random graph models.
\newblock \emph{Sociological Methodology}, 36\penalty0 (1):\penalty0 99--153,
  2006.

\bibitem[Strauss and Ikeda(1990)]{strauss1990pes}
David Strauss and Michael Ikeda.
\newblock Pseudolikelihood estimation for social networks.
\newblock \emph{Journal of the American Statistical Association}, 85\penalty0
  (409):\penalty0 204--212, 1990.
\newblock ISSN 0162-1459.

\bibitem[Thomas and Blitzstein(2011)]{thomas2011vtt}
Andrew~C. Thomas and Joseph~K. Blitzstein.
\newblock Valued ties tell fewer lies: {Why} not to dichotomize network edges
  with thresholds.
\newblock January 2011.
\newblock URL \url{http://arxiv.org/abs/1101.0788}.

\bibitem[van Duijn et~al.(2004)van Duijn, Snijders, and Zijlstra]{duijn2004re}
Marijtje A.~J. van Duijn, Tom A.~B. Snijders, and Bonne J.~H. Zijlstra.
\newblock $p_2$: a random effects model with covariates for directed graphs.
\newblock \emph{Statistica Neerlandica}, 58\penalty0 (2):\penalty0 234--254,
  2004.

\bibitem[Wasserman(1980)]{wasserman1980asn}
Stanley Wasserman.
\newblock Analyzing social networks as stochastic processes.
\newblock \emph{Journal of the American Statistical Association}, 75\penalty0
  (370):\penalty0 280--294, June 1980.
\newblock ISSN 0162-1459.

\bibitem[Wasserman and Pattison(1996)]{wasserman1996lml}
Stanley~S. Wasserman and Philippa Pattison.
\newblock Logit models and logistic regressions for social networks: {I.} {An}
  introduction to {Markov} graphs and $p^*$.
\newblock \emph{Psychometrika}, 61\penalty0 (3):\penalty0 401--425, 1996.
\newblock ISSN 0033-3123.

\bibitem[Webster and Driskell(1983)]{webster.driskell:ajs:1983}
Murray Webster and James~E. Driskell.
\newblock Beauty as status.
\newblock \emph{American Journal of Sociology}, 89:\penalty0 140--165, 1983.

\bibitem[White et~al.(1976)White, Boorman, and Breiger]{white1976ssm}
Harrison~C. White, Scott~A. Boorman, and Ronald~L. Breiger.
\newblock Social structure from multiple networks. {I.} {Blockmodels} of roles
  and positions.
\newblock \emph{The American Journal of Sociology}, 81\penalty0 (4):\penalty0
  730--780, 1976.
\newblock ISSN 00029602.

\end{thebibliography}

\newpage

\renewcommand{\theequation}{A\arabic{equation}}
\setcounter{equation}{0}
\renewcommand{\thesection}{\Alph{subsection}}
\setcounter{section}{0}

\section*{\label{app:rank-MH} Appendix: \newtxt{Overview of Monte Carlo MLE and }a simple sampling algorithm for a complete ranking ERGM}
\newtxt{In this appendix, we review the basic algorithm of
\citet{geyer1992cmc}, its application to ERGMs for dichotomous data by
\citet{hunter2006ice}, and its extension to valued ERGMs by
\citet{krivitsky2012erg}, and outline the Markov chain Monte Carlo
sampling algorithm that it requires for rank data.}

\newtxt{The challenge in fitting the model specified by \eqref{eq:rankergm} is
that the likelihood contains an intractable normalizing constant
\eqref{eq:rankergmc}, which is a summation over a large, finite
sample space of cardinality $\en\{\}{(\nactors-1)!}^\nactors$. This
summation can be approximated using importance sampling Monte Carlo
integration. In practice, this is highly inefficient. For the purposes
of likelihood maximization, however, it suffices to approximate the
ratio of the normalizing constants, i.e.,
$\ceg(\paramv')/\ceg(\paramv)$ for some two parameter configurations
$\paramv'$ and $\paramv$: maximizing
\begin{equation}\frac{\Prob_{\paramv';\genstats}(\Yy)}{\Pteg(\Yy)}=\left.\frac{\exp\en\{\}{\innerprod{\paramv'}{\genstats(\yv)}}}{\ceg(\paramv')}\middle/\frac{\exp\en\{\}{\innerprod{\paramv}{\genstats(\yv)}}}{\ceg(\paramv)}\right.=\left.\frac{\exp\en\{\}{\innerprod{\paramv'}{\genstats(\yv)}}}{\exp\en\{\}{\innerprod{\paramv}{\genstats(\yv)}}} \middle/ \frac{\ceg(\paramv')}{\ceg(\paramv)}\right.\label{eq:lr}\end{equation}
with respect to $\paramv'$ will produce the MLE as well as maximizing
$\Prob_{\paramv';\genstats}(\Yy)$ itself would.}

\newtxt{In estimating this ratio, in particular, for a given ERGM
configuration $\paramv'$, it makes sense to use a proposal
distribution from the same family, with a similar $\paramv$, which,
for an exponential family in particular, simplifies as follows:
\begin{align}
  \frac{\ceg(\paramv')}{\ceg(\paramv)}&=\frac{\sum_\ynetsY \exp\en\{\}{\innerprod{\paramv'}{\genstats(\yv)}}}{\ceg(\paramv)}
  =\frac{\sum_\ynetsY \exp\en\{\}{\innerprod{\left(\paramv'-\paramv\right)}{\genstats(\yv)}}\exp\en\{\}{\innerprod{\paramv}{\genstats(\yv)}}}{\ceg(\paramv)}\notag\\
  &=\sum_\ynetsY \exp\en\{\}{\innerprod{\left(\paramv'-\paramv\right)}{\genstats(\yv)}}\frac{\exp\en\{\}{\innerprod{\paramv}{\genstats(\yv)}}}{\ceg(\paramv)}.\label{eq:mcmle1}
\end{align}
Now, $\exp\en\{\}{\innerprod{\paramv}{\genstats(\yv)}}/\ceg(\paramv)$
is just $\Pteg(\Yy)$, making \eqref{eq:mcmle1} an expectation of $\exp\en\{\}{\innerprod{\left(\paramv'-\paramv\right)}{\genstats(\Yv)}}$ under $\Yv\sim \Pteg(\cdot)$, so the ratio can be approximated simply with
\begin{equation}
\frac{\ceg(\paramv')}{\ceg(\paramv)}=\Eteg\EN[]{\exp\en\{\}{\innerprod{\left(\paramv'-\paramv\right)}{\genstats(\Yv)}}}\approx\frac{1}{S}\sum_{s=1}^S\exp\en\{\}{\innerprod{\left(\paramv'-\paramv\right)}{\genstats(\Yv^{(s)})}},\label{eq:mcmle-sum}
\end{equation}
given a sample $\Yv^{(1)},\dotsc,\Yv^{(S)}$ drawn from $\Pteg(\cdot)$
for a given guess $\paramv$.  The accuracy of approximation
\eqref{eq:mcmle-sum} decreases as $\paramv'$ draws father from
$\paramv$, because the greater the difference between them, the
greater the variance of the quantity in the exponent of
\eqref{eq:mcmle-sum}, so, in practice, a starting guess
$\paramv^{(0)}$ is selected, sampled from, used to maximize
\eqref{eq:lr} to produce a new guess $\paramv^{(1)}$, which is in turn
used to draw a new sample, and so on until convergence.}

\newtxt{Notice that \eqref{eq:mcmle-sum} only depends on $\Yv^{(s)}$ through
its vector of sufficient statistics $\genstats(\Yv^{(s)})$. This means that once the
sample is drawn, the maximization is agnostic to nature of the
data or the structure of the sample space $\netsY$, depending only on
the sufficient statistics. In practice, this reduces its storage
requirements and greatly simplifies convergence and other diagnostics,
and it means that developments such as those of \citet{HuHu12i} to
improve stability and accuracy of the estimation can be applied
directly. We use them here as well.}

\newtxt{To obtain such a sample, we use a Metropolis sampling algorithm (Algorithm~\ref{alg:rank-MH}). At every iteration, an ego is chosen at
random and a symmetric proposal to swap the rankings of two of its
alters (also chosen at random) is made.} Any permutation of
$\nactors-1$ alters for each ego can be reached from any other
permutation in at most $\nactors-2$ swaps, and the sample space is
finite, guaranteeing ergodicity.

\begin{algorithm}
\caption{\label{alg:rank-MH}Sampling from a complete rank ERGM}
\begin{flushleft}
  \begin{description}[partopsep=0pt,topsep=0pt,parsep=0pt,itemsep=0pt]
  \item[Let:]\
    \begin{description}[partopsep=0pt,topsep=0pt,parsep=0pt,itemsep=0pt]
    \item[$\RandomChoose(A)$] return a random element of a set $A$
    \item[$\Uniform(a,b)$] return a random draw from the
      $\Uniform(a,b)$ distribution
    \end{description}
  \end{description}
\end{flushleft}
\begin{algorithmic}[1]
  \REQUIRE $\yv^{(0)}\in\netsY$, $S$ sufficiently large, $\actors$, $\genstats(\cdot)$\newtxt{, $\paramv$}
  \ENSURE a draw from a complete ranking ERGM $\Pteg(\cdot)$
  \FOR{$s \gets \fromthru{1}{S}$}
  \STATE $i \gets \RandomChoose(\actors)$ \COMMENT{Select an ego at random.}
  \STATE $j \gets \RandomChoose(\actorsnot{i})$ \COMMENT{Propose one alter.}
  \STATE $j' \gets \RandomChoose(\actorsnot{i,j})$ \COMMENT{Propose another alter.}
  \STATE $\yv^*\gets \egoswapr{\left(\yv^{(s-1)}\right)}{i}{j}{j'}$ \COMMENT{Propose a swap.}
  \STATE $r\gets \exp\EN[]{\innerprod{\paramv}{\en\{\}{\genstats(\yv^*)-\genstats(\yv^{(s-1)})}}}$
  \STATE $u\gets \Uniform(0,1)$
  \IF{$u<r$}
  \STATE $\yv^{(s)}\gets\yv^*$ \COMMENT{Accept the proposal.}
  \ELSE
  \STATE $\yv^{(s)}\gets\yv^{(s-1)}$ \COMMENT{Reject the proposal.}
  \ENDIF
  \ENDFOR
  \RETURN $\yv^{(S)}$
\end{algorithmic}
\end{algorithm}

\end{document}